\documentclass[pra,showpacs,twocolumn,superscriptaddress,floatfix,aps]{revtex4-2}
\usepackage{amsmath,amssymb,bm}
\usepackage{dsfont}
\usepackage{comment}
\usepackage{color}
\usepackage{float}
\usepackage{appendix}
\usepackage{soul}
\usepackage{hyperref}
\usepackage{braket}

\usepackage[normalem]{ulem}
\hypersetup{colorlinks,
	linkcolor=blue,%
	citecolor=blue,%
	urlcolor=blue}
\usepackage[T1]{fontenc}

\newcommand{\me}[1]{\textcolor{black}{#1}}
\usepackage{graphicx}
\usepackage{xcolor}
\usepackage{bbding}
\usepackage{tikz}
\usepackage{xcolor}
\usepackage{pifont}
\usepackage[utf8]{inputenc}
\begin{document}

\title{Excitation spectrum of vortex-lattice modes in {a} rotating condensate with {a} density-dependent gauge {potential}}
\author{Rony Boral}
\affiliation{Department of Physics, Indian Institute of Technology, Guwahati 781039, Assam, India} 

\author{Swarup K. Sarkar}
\affiliation{Department of Physics, Indian Institute of Technology, Guwahati 781039, Assam, India} 

\author{Matthew Edmonds}
\affiliation{ARC Centre of Excellence in Future Low-Energy Electronics Technologies, School of Mathematics and Physics, University of Queensland, St Lucia, QLD 4072, Australia}
\affiliation{Department of Physics and Research and Education Center for Natural Sciences, Keio University, Hiyoshi 4-1-1, Yokohama, Kanagawa 223-8521, Japan} 

\author{Paulsamy Muruganandam}
\affiliation{Department of Physics, Bharathidasan University, Tiruchirappalli 620024, Tamilnadu, India}
\affiliation{Department of Medical Physics, Bharathidasan University, Tiruchirappalli 620024, Tamilnadu, India}

\author{Pankaj Kumar Mishra}
\affiliation{Department of Physics, Indian Institute of Technology, Guwahati 781039, Assam, India}

\date{\today}

\begin{abstract}

We investigate the collective excitation spectrum of a quasi-2D Bose-Einstein condensate trapped in a harmonic confinement with nonlinear rotation induced by a density-dependent gauge field. Using {a} Bogoliubov–de Gennes(BdG) analysis, we show that the {dipole mode} frequency depends strongly on the nonlinear interaction strength, violating Kohn's theorem. Further utilizing the variational analysis, we derive analytical expressions for the dipole and breathing modes, which suggests a strong dependence of the condensate's width on the nonlinear rotation resulting from the density-dependent gauge potential. We identify four different vortex displacement modes - namely Tkachenko, circular, quadratic, and rational - whose frequencies are sensitive to the nonlinear rotation. In addition to the numerical analysis, we also derive an analytical expression for the Tkachenko mode frequency using \me{a} Hydrodynamic approach that agrees well with the frequencies obtained by the Fourier analysis of the transverse and longitudinal vortex dynamics induced by a Gaussian perturbation as well as the frequencies from \me{the} BdG excitation spectrum. Our findings also reveal that the excitation spectrum remain symmetric \me{around the} angular quantum number $l=0$, with modified energy splitting between $l$ and $-l$ \me{as} the nonlinear rotation \me{changes} from negative to positive \me{values}. Finally, we demonstrate that the surface mode excitation \me{frequency} increases (decreases) with \me{an} increase in the positive (negative) nonlinear rotation \me{strength}. 


\end{abstract}

\maketitle
\section{Introduction}
Over the past few decades, research on topological excitations in various quantum liquids \me{has attracted} significant interest~\cite{Anderson:2007, Leggatt:2006}, encompassing areas such as superfluid helium~\cite{Donnelly:1991}, liquid crystals~\cite{Malomed:2005}, quantum magnets~\cite{Zapf:2014}, and topological systems~\cite{Sonin:1987}. The discovery of vortex lattice phase in rotating Bose-Einstein condensates (BECs) has opened up new possibilities for exploring different phases and excitations of topological vortex states in a more controlled setting~\cite{Fetter:2009}. These developments have garnered considerable interest from the ultracold atom community, driven by the variety of excitation modes, including Tkachenko waves, which arise from vortex displacement modes observed in both superfluid helium and BECs~\cite{Sonin:2014}.

Several works have focused on analyzing the underlying mechanism behind the excitation of vortex lattice displacement \me{modes} in rotating BECs with short range \me{interactions}. For instance, Coddington \textit{et al.}~\cite{Coddington:2003} experimentally observed the Tkachenko mode by focusing a resonant laser beam at the center of the condensate. Subsequently, there \me{were} a series of theoretical approaches that have been adopted to develop an adequate hydrodynamic description of the Tkachenko mode~\cite{Anglin:2002, Baym:2003, Sonin:2005, Cozzini:2005, Baksmaty:2004, Mizushima:2004}. In particular, Baym~\cite{Baym:2003} analytically derived the Tkachenko mode within the framework of the continuum approximation and reported that the mode frequency approaches zero in the centrifugal limit, indicating the influence of finite compressibility on the mode's dynamics. The mode frequency remains linear in the stiff mode region, while it shows a quadratic nature in the soft range. Ke\c{c}eli and Oktel~\cite{Kecceli:2006} extended the study of Tkachenko mode for two-component BECs and \me{predicted} the appearance of optical and acoustic modes.

The low-lying collective \me{excitations} of rotating BECs reveal important \me{insights} into the stability of different \me{topological} phases of vortex lattice states of rotating BECs. Collective excitations involving vortices or vortex lattices have been studied both in quasi-2D and 3D~\cite{Simula:2010, Simula:2013jcm, Simula:2013}. Smith \textit{et al.}~\cite{Smith:2004} experimentally observed a gyroscopic tilting mode of a vortex array, which belongs to the common Kelvin-Tkachenko mode branch. Following this, Simula~\cite{Simula:2013} carried out a numerical analysis of the collective-mode spectra using Bogoliubov–de Gennes (BdG) theory in three-dimensional rotating BEC within an oblate harmonic trap and reported a comprehensive analysis of the motion of the vortices corresponding to the Kelvin-Tkachenko waves. In other directions, the stability and excitations of more complex systems like dipolar condensates with a vortex lattice have \me{been} shown to host more complex excitations which lead to the tilting and bending of the vortex core, responsible for the local instability of the BECs~\cite{Wilson:2009}. Jia \textit{et al.}~\cite{Jia:2018} subsequently extended the analysis of low-lying excitations of the vortex lattice in the presence of the anisotropic dipolar interaction and reported that in the limit of fast rotation and strong \me{dipole-dipole interactions} vortices form stripe lattices which results \me{in} the appearance of \me{a} shearing motion among the low lying Tkachenko modes, a phenomenon absent in the conventional vortex lattice with isotropic interactions.

The {observation} of artificial electromagnetism in ultracold gas systems has attracted immense interest among researchers due to its manifestation in a variety of novel phenomena~\cite{Dalibard:2011, Goldman:2014}. This induced electromagnetism gives rise to two types of gauge potentials, namely static and quasi-dynamic. In the static case, where feedback between the light and the matter wave is absent, they have been experimentally shown to exhibit various interesting phenomena, such as, including orbital magnetism~\cite{Lin:2009,Lin:2009nature,Lin:2011}, spin-orbit coupling \me{with bosons}~\cite{Lin:2011spin} \me{and fermions}~\cite{Cheuk:2012spin, Wang:2012spin} and spin-angular-momentum coupling with bosons~\cite{Chen:2018spin, Chen:2018rotating, Zhang:2019ground}. 
In the quasi-dynamical case however, the interaction between the gauge potential and the quantum state leads to the appearance of time-dependent feedback in the form of a density-dependent gauge potential, which has been \me{theoretically} explored \me{in} continuum systems~\cite{Edmonds:2013simulating} and lattice-based \me{geometries}~\cite{Keilmann:2011statistically, Greschner:2015density, Jamotte:2022strain}. Apart from this, experimental realizations of density-dependent magnetism have also been achieved in lattice-based systems for both bosonic~\cite{Clark:2018} and fermionic~\cite{Gorg:2019} systems as well as in ensembles of Rydberg atoms~\cite{Lienhard:2020}. Complementary to the lattice implementations, density-dependent gauge theories have also been realised in the continuum observing unusual topoligcal effects associated with domain walls \cite{Yao:2022} and chiral solitons \cite{Frolian:2022,Chrisholm:2022}.

Theoretical models have played an {important} role in {building} understanding {of the phenomenology associated with synthetic gauge theories. Here interest has focused on understanding the properties of this unusual nonlinear system, such as the solitary wave solutions} ~\cite{Aglietti:1996,Harikumar:1998,Dingwall:2018, Dingwall:2019,Ohberg:2019,Bhat:2021,Jia:2022,Xu:2023,Arazo:2023,Arazo:2024,Gao:2023,Ohberg:2024} and further their transport mechanisms in double-well~\cite{Edmonds:2013}{, optical-lattice \cite{Zhang:2023}} and harmonic potentials~\cite{Saleh:2018,Zhou:2024}. Theoretical exploration on collective excitations indicate that the presence of gauge potentials leads to the violation of Kohn's theorem. {Previously}, it has been {analytically} and numerically demonstrated that density-dependent gauge potentials induce {irregular dynamics \cite{Edmonds:2015} and} chaotic collective behavior in {two-dimensional} BECs~\cite{Zheng:2015}. The density-dependent gauge potential {leads to} nonlinear rotation in the system~\cite{Butera:2015,Butera:2016,Butera:2017,Correggi:2019} ultimately responsible for the formation of non-Abrikosov vortex lattices and ring vortex like arrangements~\cite{Edmonds:2020}, {while subsequent studies have focused on understanding the} critical rotation frequency for vortex nucleation~\cite{Bhat:2023} {and the appearance of ghost vortices \cite{Bhat:2024}}.

The exploration of vortex lattice displacement modes under the influence of the density-dependent gauge potential remains {an open question}. Building upon the formalism developed for the density-dependent gauge {potentials} in rotating condensates, we present a detailed analysis of the vortex displacement modes and excitations for rotating {condensates} in {the} presence of nonlinear rotation induced by the gauge potential by solving the Bogoliubov-de Gennes (BdG) equations. Our results are supported by the time evolution of the expectation values of physical observables, which are analyzed after introducing an appropriately chosen perturbation to the time-dependent Gross-Pitaevskii equation. 

Through {a} variational analysis, we analytically calculate the frequencies of the {dipole} and breathing modes. In addition, we derive the frequency of the vortex displacement modes associated with the Tkachenko wave using hydrodynamic continuum theory. This analytical approach allows us to critically compare our findings with those obtained from the BdG analysis and the time evolution of physical observables. Furthermore, we show that the energy splitting between the angular quantum numbers $+l$ and $-l$ is induced by the nonlinear rotation resulting from the density-dependent gauge potential.

The structure of {the} paper is as follows. In Sec.~\ref{sec:model}, we introduce the \me{mean-field} Gross-Pitaevskii model incorporating the density-dependent gauge potential and further provide the simulation details. Sec.~\ref{sec:collectiveExcitation} presents the effect of the density-dependent gauge potential on the ground state and the low-lying excitation spectrum. Section~\ref{sec:TimePert} presents the real-time dynamics of the perturbed ground state, illustrating the {dipole mode} oscillation frequency and performing variational calculations to derive the equations of motion for the center of mass. Section~\ref{sec:vort-disModel} focuses on the effect of the density-dependent gauge potential on the vortex displacement mode, providing an analytical expression for the Tkachenko mode frequency using hydrodynamic continuum theory. In Sec.~\ref{sec:conclusion}, we summarise our findings.

\section{Theoretical Model} \label{sec:model} 
We consider {a} weakly interacting Bose-Einstein condensate (BEC) composed of $N$ two-level atoms. These atoms are coupled through a coherent light-matter interaction driven by an applied laser field. The Hamiltonian of this system, utilizing the rotating wave approximation {is} ~\cite{Dalibard:2011}:
\begin{align}\label{eq:ham}
 \hat{\mathcal{H}} = \left(\frac{\mathbf{\hat{p}}^2}{2m}+ V(\mathbf{r})\right)\otimes\mathds{1} + \hat{\mathcal{H}}_{\text{int}} + \hat{\mathcal{U}}_{\text{MF}},
\end{align}
here $\mathbf{\hat{p}}$ is the momentum operator, $V(\mathbf{r}) = m(\omega^2_{x}x^2 + \omega^2_{y}y^2 + \omega^2_{z}z^2)/2$ is the trapping potential and $\mathds{1}$ represents the $2\times 2$ {identity} matrix. The mean-field interactions are described by $\hat{\mathcal{H}}_{\text{int}} = (1/2)\text{diag}[\Delta_{1},\Delta_{2}]$, where $\Delta_{i}=g_{ii}\vert \Psi_{i}\vert ^2 + g_{ij}\vert \Psi_{j}\vert ^2$ and $g_{ij} = 4\pi \hbar^2 a_{ij}/m$ with $a_{ij}$ being the scattering lengths for collisions between atoms in internal states $i$ and $j$ ($i, j = 1, 2$)~\cite{Goldman:2014}. The light-matter interaction, denoted as $\hat{\mathcal{U}}_{\text{MF}}$, can be defined as ~\cite{Goldman:2014} 
\begin{align}\label{eq:lm}
\hat{\mathcal{U}}_{\text{MF}} = \frac{\hbar \Omega_{r}}{2}
\begin{pmatrix}
 \cos\theta\left(\mathbf{r}\right)& e^{-i\phi\left(\mathbf{r}\right)}\sin \theta\left(\mathbf{r}\right) \\
 e^{i\phi\left(\mathbf{r}\right)}\sin \theta\left(\mathbf{r}\right) & -\cos \theta\left(\mathbf{r}\right)
 \end{pmatrix}.
\end{align}
Here $\Omega_{r}$ represents the Rabi frequency, $\theta\left(\mathbf{r}\right)$ denotes the mixing angle, and $\phi\left(\mathbf{r}\right)$ indicates the phase of the incident laser beam. {In \me{order} to introduce interactions into the dressed states,} perturbed dressed states can be {defined} using perturbation theory~\cite{Edmonds:2013simulating}
\begin{align}
 \ket{\Psi_{\pm}} = \ket{\pm} \pm \frac{\Delta_{d}}{\hbar \Omega_{r}}\ket{\mp}
\end{align}
with $\ket{\pm}$ representing the unperturbed eigenstates of $\hat{U}_{\text{MF}}$, and $\Delta_{d} = \text{sin}\left(\theta/2\right) 
\text{cos}\left(\theta/2\right) \left(\Delta_{1}-\Delta_{2}\right)/2$
denotes the mean-field detuning. The qualitative physics remains unaffected by {the choice of dressed} state, {hence we project into the $+$ state, giving}
\begin{align}\label{eq:ham+}
 \mathit{\hat{H}}_{+} = \frac{\left(\mathbf{\hat{p}}-\mathbf{A}_{+}\right)^2}{2m} + W_{+} + \frac{\hbar \Omega_{r}}{2} + \Delta_{+} + V (\mathbf{r}),
\end{align}
where $\mathbf{A}_{+} = i\hbar\braket{\Psi_{+} \vert \Psi_{+}}$, and $W_{+} = \hbar^{2} \vert \braket{\Psi_{+} \vert \nabla \Psi_{-}} \vert^{2} / 2m$ are the vector gauge potential and scalar potential, respectively. $\Delta_{+} = \big(\Delta_{1} \cos^{2}(\theta/2) + \Delta_{2} \sin^{2}(\theta/2)\big)/2$ is the dressed mean-field \me{detuning}. The vector potential \me{is} denoted as $\mathbf{A}_{+}$, and the scalar potential, denoted as $W_{+}$, are defined as follows:
\begin{subequations}\label{eqn:awpot}
	\begin{align}\label{eq:avec}
		{\bf A}_+=&-\frac{\hbar}{2}(1-\cos\theta)\nabla\phi+\frac{\Delta_{\rm d}}{\Omega_r}\nabla\phi\sin\theta,\\ \notag 
		W_+=&\frac{\hbar^2}{8m}(\nabla\theta)^2+\frac{\hbar}{8m}\sin^2\theta(\nabla\phi)^2\\+&\frac{\hbar}{2m}\frac{\Delta_{\rm d}}{\Omega_r}\sin\theta\cos\theta(\nabla\phi)^2-\hbar\nabla\theta\cdot\nabla\frac{\Delta_{\rm d}}{\Omega_r}.\label{eq:wsca}
	\end{align}
\end{subequations}

The extremization of the energy functional~$\mathcal{E}=\braket{\Psi_{+}\vert \hat{H}_{+}\vert \Psi_{+}}$ then yields the generalized Gross-Pitaevskii (GP) equation for $\Psi_{+}$~\cite{Butera:2017}:
\begin{align}\label{eq:ggpe}\notag 
 \mathrm i\hbar &\frac{\partial\Psi_{+}}{\partial t} = \left[ \frac{\left(\mathbf{\hat{p}}-\mathbf{A}_{+}\right)^2}{2m} + \mathbf{a}_{1}\cdot\mathbf{J}\right]\Psi_{+} +\\ \notag 
 & \left[V(\mathbf{r})+\frac{\hbar\Omega_{r}}{2}+2\Delta_{+}+W_{+}\right]\Psi_{+} +\\
 & \left[n_{+} \left(\frac{\partial W_{+}}{\partial \Psi^{*}_{+}} - \boldsymbol\nabla \cdot \frac{\partial W_{+}}{\partial \boldsymbol\nabla\Psi^{*}_{+}} \right) - \frac{\partial W_{+}}{\partial \boldsymbol\nabla\Psi^{*}_{+}} \cdot \boldsymbol\nabla n_{+}\right].
\end{align}
Here $n_{+} = \vert \Psi_{+}(\mathbf{r},t)\vert ^{2}$ is the atomic density in the
$\ket{\Psi_{+}}$ state, $\mathbf{a}_{1} = \boldsymbol\nabla\phi \Delta_{d}\text{sin}\theta/
n_{+}\Omega_{r}$ is the coupling strength employed by 
the gauge {potential} and 
\begin{align}\label{eq:nlc}
 \mathbf{J} {=} \frac{\hbar}{2m \mathrm i}\left[\Psi_{+}\left(\boldsymbol\nabla {+} \frac{\mathrm i}{\hbar}\mathbf{A}_{+}\right)\Psi^{*}_{+}{-}\Psi^{*}_{+}\left(\boldsymbol\nabla{-}\frac{\mathrm i}{\hbar}\mathbf{A}_{+}\right)\Psi_{+}\right]
\end{align} 
is the nonlinear current. The expressions for the vector and scalar potentials are simplified by expanding Eqs.~(\ref{eq:avec}) and (\ref{eq:wsca}) up to first order in terms of $\theta$ and $\epsilon$. Here $\theta=\Omega_{r}/\Delta$ represents the ratio of Rabi frequency to laser detuning, and $\epsilon = n_{+}(\mathbf{r})(g_{11}-g_{12})/4\hbar\Delta$ accounts for the collisional and coherent interactions. \me{Then, Eqs.~\eqref{eqn:awpot} simplify to} 
\begin{align}\label{eq:vpot}
\mathbf{A}_{+} = & -\frac{\hbar \theta^{2}}{4}\left(1-4\epsilon\right)\boldsymbol\nabla\phi, \\
\label{eq:spot}
\mathit{W}_{+}= &\frac{\hbar^{2}}{2}\bigg[\frac{\left(\boldsymbol\nabla\theta\right)^2
 \left(1-4\epsilon\right)+\theta^2\left(1+4\epsilon\right)
 \left(\boldsymbol\nabla\phi\right)^2}{4m} \notag \\
 & - \boldsymbol\nabla\theta^2 \cdot\boldsymbol\nabla\epsilon\bigg]. 
\end{align} 
To construct a generalized Gross-Pitaevskii equation, we consider a spatially dependent Rabi coupling defined by $\Omega_{\text{r}} = \kappa_{0} r$, where $r$ is the radial distance and $\kappa_{0}$ is a constant. Additionally, we define the phase as $\phi = l \varphi$, with $l$ representing the angular momentum and $\varphi$ being the polar angle of the incident laser beam. By substituting Eqs. (\ref{eq:vpot}) and (\ref{eq:spot}) into Eq. \ref{eq:ggpe}, we derive the simplified three-dimensional GP equation:
\begin{align}\label{eq:gpe3d}
 \mathrm i \hbar \frac{\partial \Psi}{\partial t} = 
 \bigg[-\frac{\hbar^2}{2m}\nabla^2 +
 V(\mathbf{r}) -\Omega_{n}(\mathbf{r},t) \hat{L}_{z} 
 + \text{g}_\text{eff} \vert \Psi\vert ^2 \bigg] \Psi.
\end{align}
In what follows we omit the ``$+$'' subscript. Here $\hat{L}_z=-i\hbar\partial/\partial\varphi$ is the \textit{z} component of the angular momentum operator. 
The density-dependent rotation, denoted as, $\Omega_{n}(\mathbf{r},t)$ takes the following form:
\begin{align}\label{eq:om_rho}
\Omega_{n}(\mathbf{r},t) = \Omega + \mathit{C} n(\mathbf{r}, t),
\end{align} 
where $\Omega$ represents the {rigid body} rotation {strength} and $n(\mathbf{r},t)= \vert \Psi(\mathbf{r},t)\vert ^2$ is the condensate density. 
Considering possible implementations, nonlinear rotation could be experimentally induced in a condensate using a laser carrying a spatially varying intensity with $l = 1$ units of orbital angular momentum. This process is similar to the creation of spin-angular-momentum-coupled Bose-Einstein condensates using Laguerre-Gaussian laser beams, which feature cylindrically varying intensity profiles and also \me{carries} an orbital angular momentum of $l = 1$~\cite{Chen:2018spin,Chen:2018rotating,Zhang:2019ground}.

To {obtain} the quasi-two-dimensional {generalized} Gross-Pitaevskii (GP) equation, we start by defining the wave function as $\Psi(\mathbf{r},t) = \psi (x,y,t)
\text{exp}(-z^{2}/2\sigma^{2}_{z})/\sqrt[4]{\pi \sigma^{2}_{z}}$ and consider that the condensate is confined in a highly anisotropic trap where $\omega_{z} \gg \omega_{\text{x,y}}$. By {integrating out the axial degree of freedom} we obtain the two-dimensional equation {of motion}
\begin{align}\label{eq:model}
\mathrm i \hbar \frac{\partial \psi}{\partial t} =
\bigg[-\frac{\hbar^2}{2m}\nabla^2 +
V(x,y) -\Omega_{n} \hat{L}_{z} 
+ \text{g}_{\text{2D}} \vert \psi\vert ^2 \bigg] \psi.
\end{align}
The nonlinear coefficients {appearing in Eq.~\eqref{eq:model} are defined as} $\text{g}_{\text{2D}} = \text{g}_{\text{eff}}/\sqrt{2\pi}\sigma_{z}$ {and $\mathit{C}\rightarrow\mathit{C}/\sqrt{2\pi}\sigma_z$}.
Here, $n(x,y,t) = |\psi(x,y,t)|^2$ denotes the {quasi-two-dimensional} density of the condensate.
By applying spatiotemporal scaling, the variables are redefined as $t = \omega^{-1}_{\perp} t, (x,y) = a_{\perp} (x^{\prime},y^{\prime}),
~\text{and}~ \psi=\sqrt{N}\psi^{\prime}/a_{\perp}$. These transformations recast Eq.~\eqref{eq:model} into the following {dimensionless} form:
\begin{align}\label{eq:dmless_eq}
\mathrm i\frac{\partial\psi}{\partial t} =
\bigg[-\frac{\nabla^2}{2} +
V(x,y) -\Omega_{n} \hat{L}_{z} + \text{g} \vert \psi\vert ^2 \bigg]\psi.
\end{align} 
In Eq.~\eqref{eq:dmless_eq}, the scaled parameters are defined as follows: $\text{g} = \text{g}_{\text{2D}} N m/\hbar^2$ and $ \Omega_{n}(x,y,t) = \Omega/\omega_{\perp} + \tilde{\mathit{C}} n(x,y,t) $ where $ \tilde{\mathit{C}} = \mathit{C}N /m\hbar\sqrt{2\pi}\sigma_{z}$. 
\begin{figure*}[!htp]
 \includegraphics[width=\linewidth]{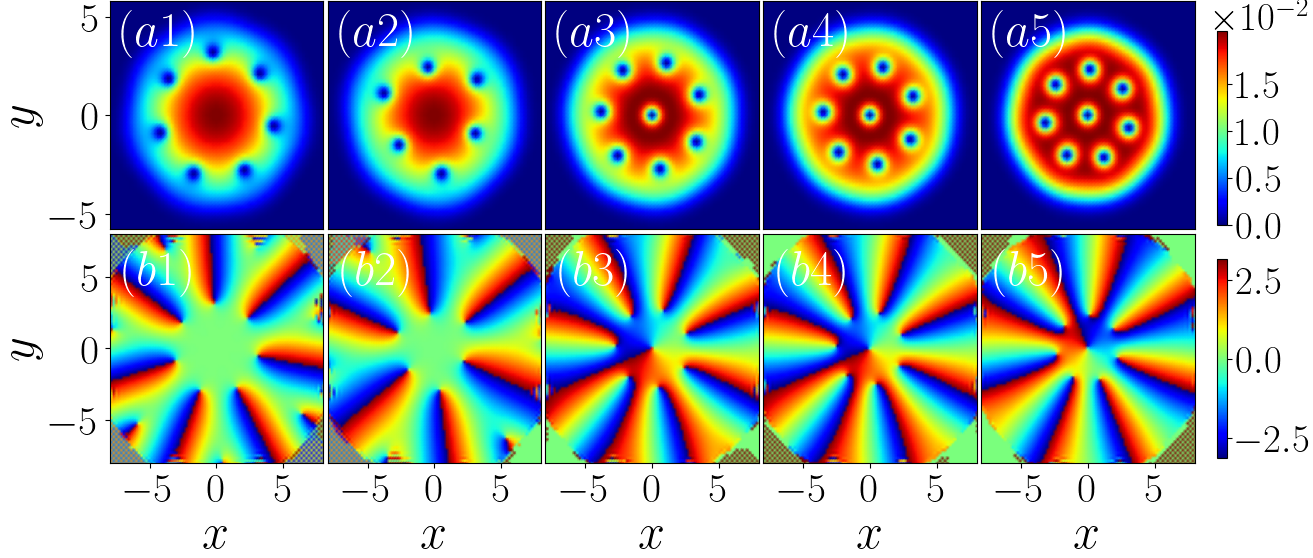}
 \caption{Pseudo-color representation of the condensate density for (a1) $\tilde{C} = -30$, (a2) $\tilde{C} = -20$, (a3) $\tilde{C} = 0$, (a4) $\tilde{C} = 10$, and (a5) $\tilde{C} = 20$ at $\Omega = 0.6$ and $\text{g}=400$. Panels (b1)–(b5) show the phase profiles of density illustrated in (a1)-(a5).}
 \label{fig:density}
 \end{figure*}

In our work we consider a condensate consisting of $N = 3 \times 10^4$ $\rm Sr$ atoms. The s-wave scattering length of the $\rm Sr$ atom is $a_s = 2.83 \, \text{nm}$. The radial trap frequency is set to $\omega_\perp = 2\pi \times 100 \, \text{Hz}$, and the trap aspect ratio is $\omega_z/\omega_\perp = 10$. These parameters give a dimensionless contact interaction strength of $g \approx 400$, which is fixed for all numerical calculations. In our simulation, we adopt spatial and temporal steps of $\Delta x = 0.07$ and $\Delta t = 0.0001$, respectively.

\section{Ground state and Collective excitation}
\label{sec:collectiveExcitation}
{Here we calculate and analyse the vortex states of the density-dependent gauge theory previously derived in Sec.~\ref{sec:model}}. Depending upon the values of $\tilde{C}$, the gauge potential induces a nonlinear rotation in the condensate, which affects the vortex lattice structure in several ways. For $\tilde{C}<0$, the vortex lattice structure shows a transition from the Abrikosov lattice structure to a ring structure. However, for $\tilde{C}>0$ the density profile exhibits a large plateau region \me{where the} vortices are mainly concentrated near the trap center~\cite{Edmonds:2020}. Following the ground state analysis, here, we aim to explore how the {nonlinear rotation alters} the low-lying collective excitation modes and {how this connects to} the {different} vortex lattice {morphologies}.

\begin{figure}[!htb]
 \includegraphics[width=1.06\linewidth]{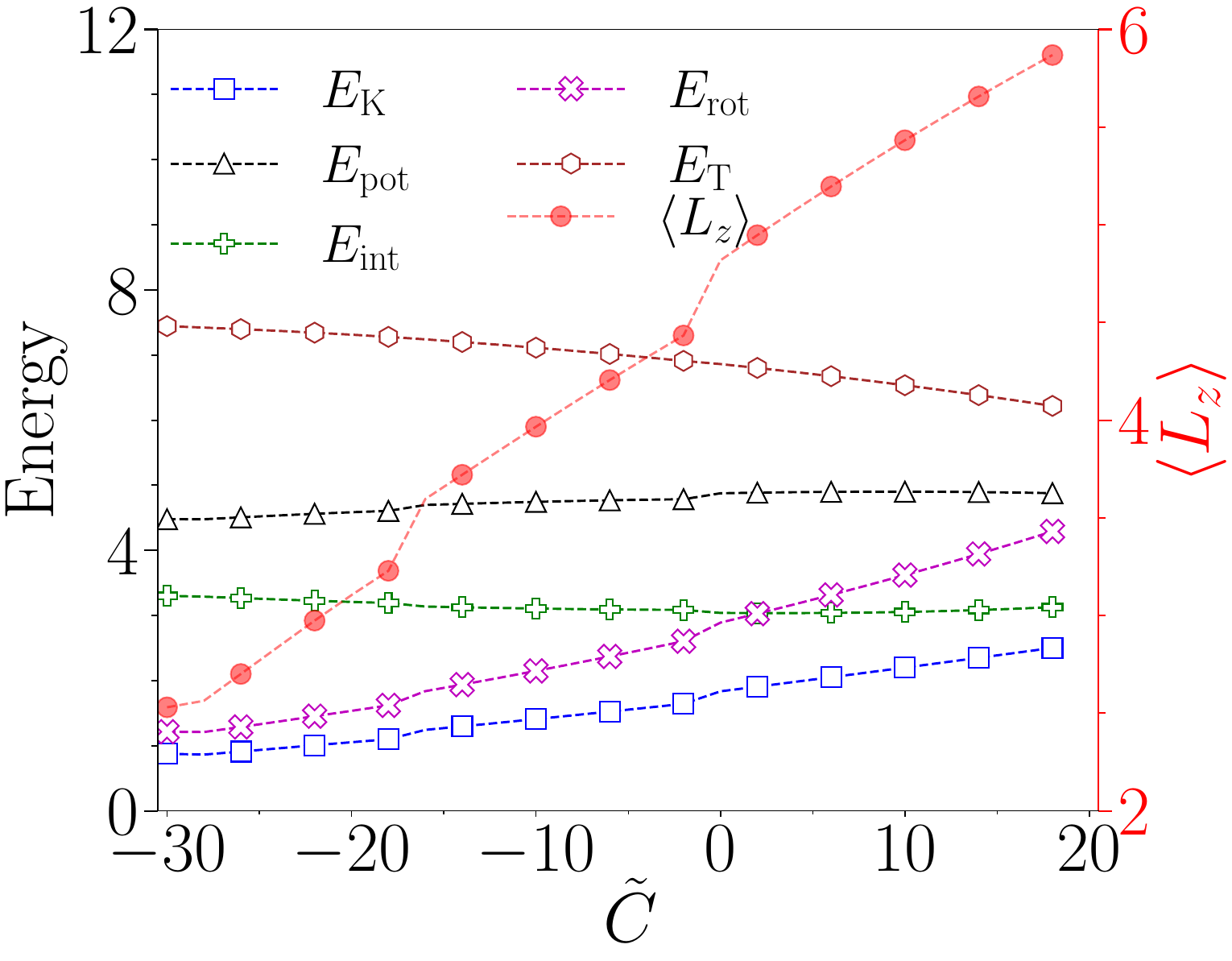}
 \caption{Variation of different \me{components of the total energy}, total energy $E_{\rm T}$, kinetic energy $E_{\rm K}$, potential energy $E_{\rm pot}$, and interaction energy $E_{\rm int}$ as a function of $\tilde{C}$. Right vertical axis shows the variation of angular momentum $\langle L_{z} \rangle$. $E_{\rm int}$ and $E_{\rm pot}$ remains almost unchanged with $\tilde{C}$, while $E_{\rm K}$ show linear increment with $\tilde{C}$. Liner increment of $\langle L_{z} \rangle$ with $\tilde{C}$ indicates in the increase in \me{the number of} vortices upon \me{increasing} $\tilde{C}$. }
 \label{fig:energy}
 \end{figure}

We begin our analysis by exploring the effect of $\tilde{C}$ on the vortex lattice structure. In Fig.~\ref{fig:density} we present the condensate density and phase profile of the condensate for different values of $\tilde{C}$ by keeping the rigid body rotation frequency fixed at $\Omega=0.6$, and the effective \me{mean-field} interaction strength as $\text{g}=400$. We observe that depending on the \me{value} of $\tilde{C}$ the gauge potential-induced nonlinear rotation can have \me{a} profound impact on the vortex lattice structure. For $\tilde{C}<0$, the vortex lattice structure \me{is} modified from the Abrikosov lattice to a ring like structure~(see Fig.~\ref{fig:density}(a1)-(a2)), while for $\tilde{C}>0$ the density profile exhibits a large plateau region, reminiscent of quantum droplet like region~\cite{Tengstrand:2019droplet} with vortices concentrated near the trap center [see Fig.~\ref{fig:density}(a4)-(a5)]. 

In order to understand the the appearance of \me{the} different vortex lattice \me{structures} in Fig.~\ref{fig:energy}, we \me{compute the different contributions to the total energy, namely the} kinetic energy $E_{\rm K}=\frac{1}{2}\int\left\vert\nabla \psi \right\vert^{2}dxdy$, potential energy $E_{\rm pot}=\frac{1}{2}\int (x^2+y^2)\left\vert\psi\right\vert^{2}dxdy$, interaction energy $E_{\rm int}=\frac{\rm g}{2}\int\left\vert\psi\right\vert^{4}dxdy$, rotational energy $E_{\rm rot}=-\int \Omega_n \psi^{*}\hat{L}_z\psi\ dxdy$ and total energy \me{$E_{\rm T}=E_{\rm K}+E_{\rm pot}+E_{\rm int}$+$E_{\rm rot}$} \me{as well as the} angular momentum $\langle\hat{L}_z \rangle$ as a function of $\tilde{C}$. 
With the variation of $\tilde{C}$, the $E_{\rm pot}$ remains almost constant while $E_{\rm int}$ exhibits slight decreasing behavior as $\tilde{C}$ increases. In contrast, $E_{\rm K}$ increases as the nonlinear rotation parameter varies from negative to positive \me{values}. This behavior arises due to the shifting of the vortices towards the periphery of the harmonic trap, \me{causing the minimization of} the energy for negative $\tilde{C}$ values. Additionally, the angular momentum $\langle \hat{L}_z \rangle$ also exhibits an increasing trend with $\tilde{C}$ due to the possibility of \me{the nucleation of} additional vortices. Similarly, $E_{\rm rot}$ also follows \me{a} similar trend \me{to the} angular momentum as $\tilde{C}$ \me{is} varied. Overall, the total energy \(E_{\rm T}\) decreases when \(\tilde{C}\) is varied.
Next we move our focus {to} investigate the collective excitations of the vortex lattice structure using the Bogoliubov–de Gennes (BdG) theory. For this, we consider the ground state $\psi_0$ is perturbed by the wave function $\delta \psi$ for which the excited wave function given by~\cite{Pitaevskii:2003,Simula:2013}:

\begin{align}
\psi(x,y,t)=\left[\psi_0(x,y)+\delta\psi(x,y,t)\right]\exp(\mathrm i\mu t) 
\label{eq:expand}
\end{align}
where $\mu$ is the chemical potential of the ground state, and the perturbation wave function $\delta\psi(x,y,t)$ is of the form:
\begin{align}
 \delta\psi(x,y,t){=}u(x,y)\exp(-i\omega t){-}v^{*}(x,y)\exp(\mathrm i\omega^{*}t),
\end{align}
where $u(x,y)$ and $v(x,y)$ are the Bogoliubov mode amplitudes and $\omega$ is the corresponding frequency. The Bogoliubov mode amplitudes follow the relation 
\begin{align}
 \int \left[\lvert u(x,y) \rvert^{2}-\lvert v(x,y) \rvert^{2}\vert \right] dxdy=1
\end{align}
By substituting Eq.~\ref{eq:expand} into Eq.~\ref{eq:dmless_eq} and equating the time dependent terms according to $\exp(-i\omega t)$ and $\exp(i\omega t)$ , respectively, we obtained the set of BdG equations
\begin{subequations}\label{eq:bdg}
	\begin{align}
		\omega u=&\left[\mathcal{L}_{1}-\tilde{C}(\rvert\psi_0\lvert^{2}L_z+\psi_{0}^{*}L_z\psi_{0})\right]u \notag \\ &
 -[\tilde{C}\psi_{0}L_z\psi_0+\text{g}\psi_{0}^{2}]v\\ 
		-\omega v=&\left[\mathcal{L}_{2}+\tilde{C}(\vert \psi_{0}\vert ^{2}L_z+\psi_{0}L_z\psi_{0}^{*})\right]v \notag \\ &
 -[\tilde{C}\psi_0^{*}L_z\psi_0^{*}-\text{g}\psi_{0}^{*2}]u
	\end{align}
\end{subequations}
Here $\mathcal{L}_{1}= -\frac{1}{2}\nabla^2+V+2\text{g}\rvert\psi_0\lvert^{2}-\mu-\Omega L_{z}$, and ~ $\mathcal{L}_{2}= -\frac{1}{2}\nabla^2+V+2\text{g}\rvert\psi_0\lvert^{2}-\mu+\Omega L_{z}$.
The set of BdG equations~\ref{eq:bdg} are solved to obtain the eigenfrequencies ~$\omega$ and eigenmodes~$u$, $v$. To visualize a normal mode, we analyze the temporal evolution of the perturbed density profile as follows:
\begin{align}
n(x,y,t)=\left[\psi_0(x,y)+p\delta\psi(x,y,t)\right]^{2} \label{eq:prtrb_den}
\end{align}
which {facilitates the study of} the excitation {spectrum in response to small perturbations to the nonlinear system}. The parameter $p \ll 1$ is small in amplitude and used to control the population of quasiparticle excitation.
\begin{figure}[!ht]
\centering
\includegraphics[width=\linewidth]{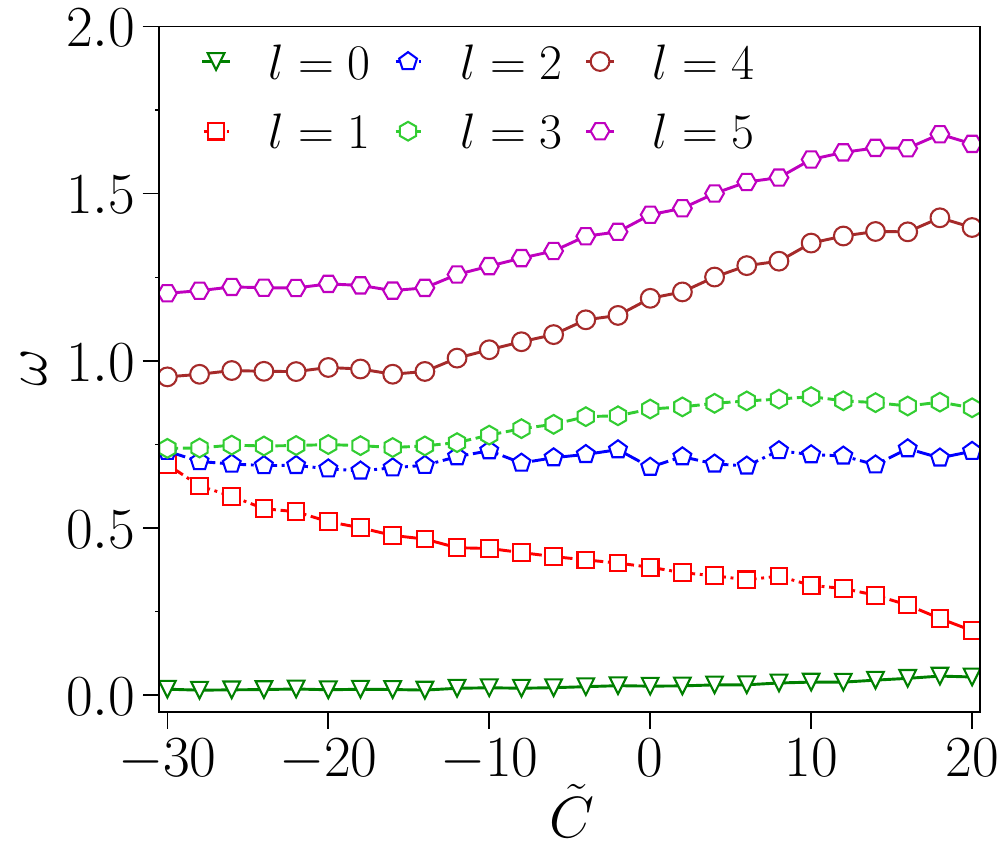}
\caption{Collective excitation spectrum for various angular quantum numbers $l$ with radial quantum number $n_r=0$ as a function of nonlinear rotation strength $\tilde{C}$ at $\Omega=0.6$ are presented. The triangles and squares denote the Tkachenko and dipole modes, respectively. The pentagon, hexagon, and circles represent the higher order surface modes.}
\label{fig:excit}
\end{figure}
\begin{figure*}[!hbt]
\centering
\includegraphics[width=\linewidth]{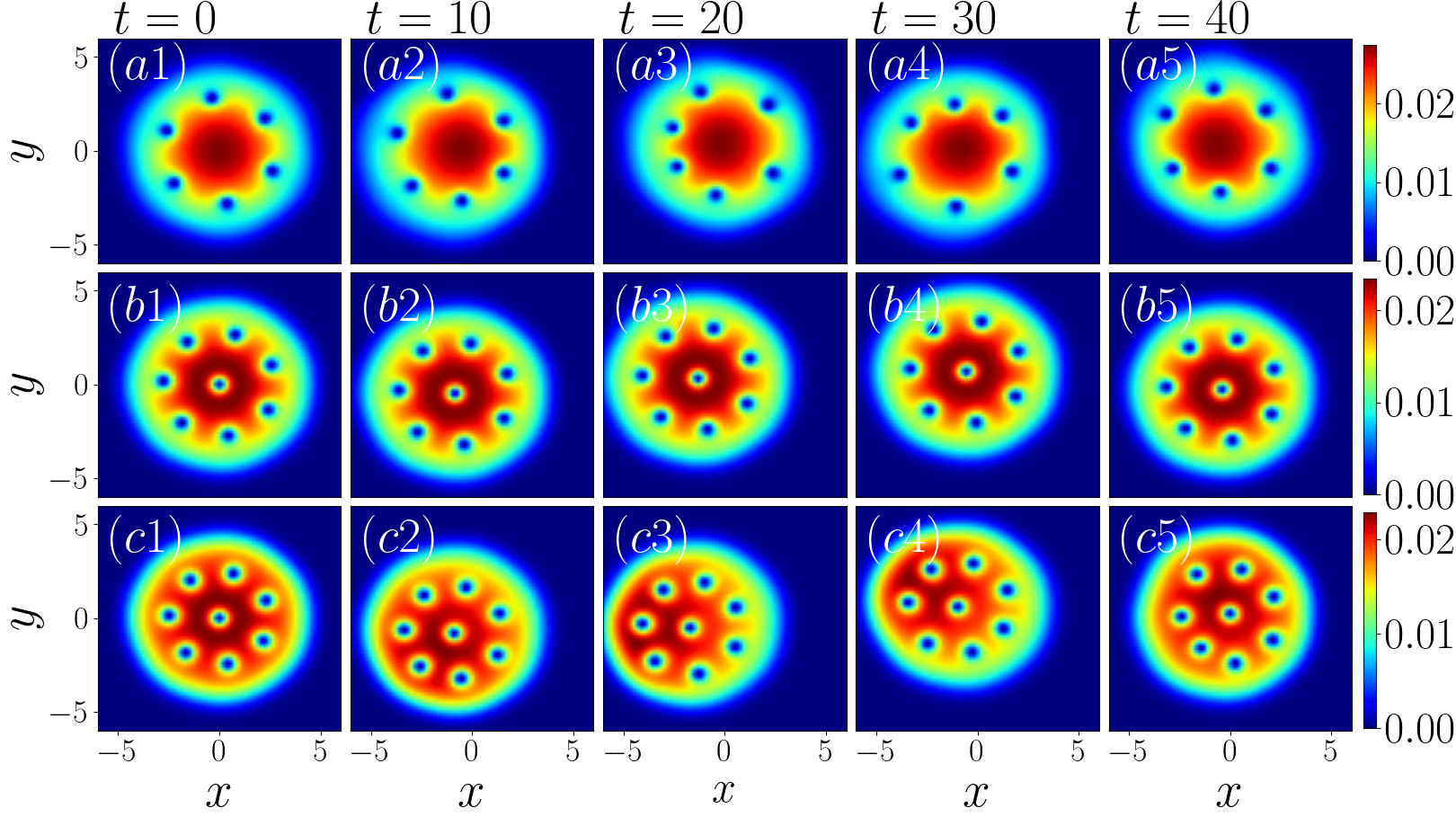}
\caption{Real space condensate density at different \me{times} portraying the vortex displacement that arises due to the application of \me{a} linear perturbation on the ground state which excites mainly the {dipole} mode for different values of $\tilde{C}$ at $\Omega=0.6$. (a1)-(a5): for fixed $\tilde{C}=-30$ and $t=(0, 10, 20, 30,40)$; (b1)-(b5): for fixed $\tilde{C}=0$ and $t=(0, 10, 20, 30,40)$; (c1)-(c5): for fixed $\tilde{C}=20$ and $t=(0, 10, 20, 30,40)$. The real-time dynamics of the condensate reveal its precessional motion around the trap center with time.}
\label{fig:time_evl_dipole}
\end{figure*}
 
Next we present the effect of density-dependent rotation on the collective excitation spectrum. To achieve this, we first calculate the excitation frequencies by numerically solving {the} BdG equations \ref{eq:bdg}. In addition to this approach, we use an alternative technique to excite the dipole mode, which is feasibly accessible in experiments. This technique defines by adding a small linear perturbation to the Hamiltonian to induce the dipole mode of the BEC. We also complement these calculations with the analytical form of the dipole frequency obtained using a variational approach. Apart from the dipole mode we also analyze the relevant vortex displacement modes by imparting a linear perturbations to the ground state of the condensate density as described in the Eq.~\ref{eq:prtrb_den}.


In Fig.~\ref{fig:excit}, we show the variation of the frequencies of {the} first \me{six} low-lying excitation modes upon as $\tilde{C}$ {is varied} for a fixed rotation frequency $\Omega=0.6$ and interaction {strength} $g=400$. The lowest frequencies~($\bigtriangledown$) corresponding to the vortex displacement excitations captures the motion of vortex cores around the equilibrium position in {the} rotating frame. Here, $l=0-1$ denotes respectively the Tkachenko, and dipole excitation, while $l\geq 2$ represents the higher order surface excitation modes~\cite{Mizushima:2004}. In the excitation spectrum, the frequencies marked by red squares ($\textcolor{red}{\Box}$) represents the dipole mode oscillations corresponding to the center of mass motion of the condensate, which exhibits decreasing trend upon increasing $\tilde{C}$. It should be noted here that as per the constraint imposed due to the Kohn theorem for condensates experiencing only rigid body rotation, usually, the dipole frequency does not show any dependency on atomic interactions. However, it can be observed that in the presence of \me{nonlinear rotation}, the dipole frequency shows a significant deviation from the Kohn theorem~\cite{Zheng:2015,Edmonds:2015}. \me{Violation of the Kohn theorem} also \me{occurs} for spin-orbit coupled \me{condensates} due to the coupling between the center of mass and the internal spin degrees of freedom, as shown in several of the works~\cite{Zhang:2012, Ozawa:2013, Chen:2012, Price:2013}. The decreasing trend of \me{the} dipole frequency can be attributed to the presence of a density-angular momentum coupling factor in the BdG equations, {per} Eqs.~\ref{eq:bdg}(a) and ~\ref{eq:bdg}(b). This coupling is responsible for the centrifugal force that is influenced by both the sign and magnitude of the nonlinear rotation parameter, $\tilde{C}$. A large value of $\tilde{C}$ breaks the symmetry of the harmonic motion of the condensate, resulting in the dipole frequency as observed in the Fig.~\ref{fig:excit}. On the other hand, we find that the higher \me{order} modes exhibit an increasing \me{frequency as} $\tilde{C}$ is varied from negative to positive values. These modes are later classified as surface excitations of the condensate based on the time evolution of the perturbed density profile, obtained using Eq.~\ref{eq:prtrb_den}. It can also be characterized by angular quantum numbers $l > 2$, which are arranged in ascending order according to their corresponding excitation frequencies.

\section{Time dependent characteristics of the collective modes}\label{sec:TimePert}

In the following section we {explore} the time-dependent characteristics of the collective modes. In addition to the linearization technique, the low-lying collective excitations, such as dipole, monopole, and quadrupole modes, can also be excited using an {additional} method, in which a time-independent linear perturbation is added to the Hamiltonian~\cite{Ueda:2010, Pitaevskii:2016, Banger:2023, Fei:2024}. In this work, our focus is to find the the {dipole} mode; thus, we only consider {a} perturbation associated with this specific mode. To excite the {dipole} mode, we consider a linear perturbative potential $V_{\rm p}(x)=\lambda x$ at $t \geq 0$ along with the harmonic potential, where $0 \me{<} \lambda \me{<} 1$. This addition modifies the Gross-Pitaevskii equation by incorporating an additional term, $V_{\rm p}(x)$. {The generalized} Gross-Pitaevskii equation Eq.~\eqref{eq:dmless_eq} {is solved} by considering the {rotating stationary} state as the initial condition to construct time-dependent physical observables, specifically focusing on the motion of the center of mass in the case of the {dipole} mode. Furthermore, the frequency of this mode is determined through Fourier transformation of the temporal evolution of the center of mass of the rotating condensate.

\begin{figure}[!htp]
 \centering
 \includegraphics[width=\linewidth]{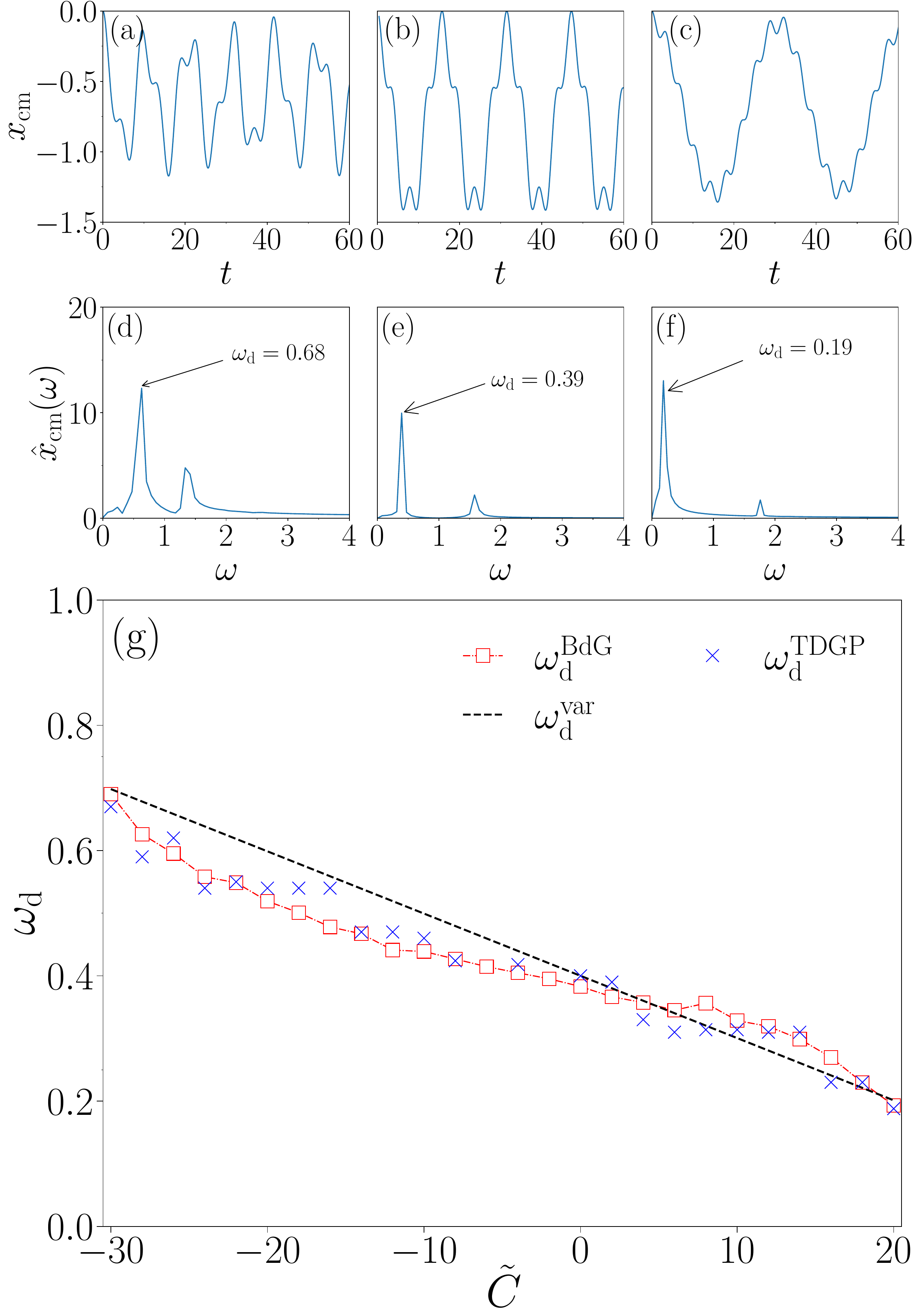}
 \caption{ Panels (a)-(c), shows the center of mass oscillation $x_{\rm cm}(t)$ of the condensate as a function of time for $\tilde{C}=-30, 0, 20$, respectively. (d)-(f) illustrates the Fourier transform of $x_{\rm cm}$. The dominant peaks observed at $\omega_{\rm d}=0.68, 39$, and $0.19$ are the {dipole} modes obtained via a linear perturbation to the condensate. (g) \me{dipole} mode frequency as a function of nonlinear rotation strength $\tilde{C}$ for $\Omega=0.6$. The black dashed line, blue crosses ~($\textcolor{blue}{\times}$), and the red squares ($\textcolor{red}{\Box}$) represent the {dipole} modes obtained using {the} variational analysis, linear perturbation to the condensate, and from {the Bogoliubov de-Gennes} equations respectively. }
 \label{fig:comparison_com}
 \end{figure}

To investigate the effect of \me{the} perturbation systematically, in Fig.~\ref{fig:time_evl_dipole} we display pseudocolor map of \me{the} density profile corresponding to the solution of Eq.~\eqref{eq:dmless_eq} at time $t=0, 20, 40$, and $60$. The analysis is conducted with the parameters $\Omega=0.6$, $\lambda=0.1$, and $\text{g}=400$. The first ((a1)-(a5)), second ((b1)-(b5)), and the third row ((c1)-(c5)) of {Fig.~\ref{fig:time_evl_dipole}} corresponds to $\tilde{C}=-30, 0$, and $20$, respectively. We find that at $t=0$, the condensate is initially positioned at the origin~(see \ref{fig:time_evl_dipole}(a1),(b1), and (c)). However, as time progresses, the perturbation causes the condensate to exhibit a dipole oscillation around the origin, {independent} of $\tilde{C}$~(see \ref{fig:time_evl_dipole}(a2)-(a5), (b2)-(b5), and (c2)-(c5)). To characterize the motion of the center of mass, we calculate the time-dependent center of mass $x_{\rm cm}=\langle x(t) \rangle$, which can be expressed as
\begin{align}
x_{\rm cm}(t)=\int\int x\left\lvert \psi(x,y,t) \right\rvert^{2} dx dy.
\end{align}
In {Fig.~\ref{fig:comparison_com}}, we present the variation of $x_{\rm cm}$~(first row) as a function of time corresponding to $\tilde{C}=-30, 0$, and $20$, demonstrating an oscillatory behavior over time~[see Fig.~\ref{fig:comparison_com}(a)-(c)]. Next, by doing the Fourier transform~[see Fig.~\ref{fig:comparison_com}(d)-(f)], we identify the dominant frequency of $x_{\rm cm}$. For $\tilde{C}=-30$, the dominant frequency is at $\omega_{\rm d}=0.68$, as the value of {nonlinear rotation strength} changes {from} ${\tilde{C}}=0$ to ${\tilde{C}}=20$, there is a noticeable shift in the dominant frequency from $\omega_{\rm d}=0.39$ to $\omega_{\rm d}=0.19$, respectively. Additionally, it is worth noting that the linear perturbation can also be given as $V_{\rm p}(y)=\lambda y$ along \me{the} $y$ direction. In \me{which} case, the dipole frequency can be obtained from $y_{\rm cm}$.

Using a combination of linearization and perturbative techniques, we observe that the {dipole} frequency {strongly} depends on the nonlinear rotation strength as shown in Fig.~\ref{fig:comparison_com}(g). To better comprehend the behavior of the {dipole} frequency we {employ a} variational approach {\cite{Pa:1997,Pa:1997b}}. The motivation behind using {this} method has two primary aspects: first, to derive an analytical expression for the frequencies of the dipole mode, and second, to establish an equation of motion that provides an accurate solution to the dynamics under nonlinear rotation.
This method has been {used to understand} both single-component~\cite{Pa:1997}, {binary \cite{Bisset:2018}} and spin-orbit coupled condensates~\cite{Chen:2012, Zhang:2012} to describe collective excitations. Following a similar approach, we consider the variational wave function as Gaussian with varying center-of-mass coordinates and width of the condensate. The normalized Gaussian ansatz \me{appropriate for a condensate confined in a} harmonic potential is given by 
\begin{align} \notag 
&\psi(x,y,t)= \frac{1}{\sqrt{\pi\sigma(t)^{2}}}\exp\Biggl[-\frac{\bigl(x-x_0(t)\bigr)^{2}}{2\sigma(t)^{2}}\\ \notag &+i\alpha_x(t)\bigl(x-x_0(t)\bigr)+i\beta_x(t)\bigl(x-x_0(t)\bigr)^{2}\\ \notag &-\frac{\bigl(y-y_0(t)\bigr)^{2}}{2\sigma(t)^{2}} 
+\mathrm i\alpha_y(t)\bigl(y-y_0(t)\bigr)\\ &+i\beta_y(t)\bigl(y-y_0(t)\bigr)^{2} \Biggr]. 
\end{align}
Where the variational parameters are \me{the} width~$\sigma(t)$, center of mass~$x_0(t)$, and $y_0(t)$, slopes $\alpha_{x,y}(t)$, and $(\rm radius \ of \ curvature)^{-1/2}$ $\beta_{x,y}(t)$.
The Euler-Lagrange \me{equations} of motion \me{are} then expressed as:
\begin{subequations}\label{eq:EOM}
\begin{align}
\ddot{x}_0 & - 2\dot{y}_0\left( \Omega + \frac{\tilde{C}}{4\pi\sigma^{2}} \right) + x_0\left[1 - \left( \Omega + \frac{\tilde{C}}{4\pi\sigma^{2}} \right)^{2} \right] \notag \\
& + \frac{\tilde{C}\dot{\sigma}}{2\pi\sigma^{3}}y_0=0 \\
\ddot{y}_0 & + 2\dot{x}_0\left( \Omega + \frac{\tilde{C}}{4\pi\sigma^{2}} \right) + y_0\left[1 - \left( \Omega + \frac{\tilde{C}}{4\pi\sigma^{2}} \right)^{2} \right] \notag \\
& - \frac{\tilde{C}\dot{\sigma}}{2\pi\sigma^{3}}x_0 = 0 \\
\ddot{\sigma} &-\frac{1}{\sigma^{3}}-\frac{\rm g}{2\pi\sigma^{3}}+\frac{\tilde{C}}{2\pi\sigma^{3}}\Biggl[(x_0\dot{y}_0-y_0\dot{x}_0) \notag \\
&+(\Omega+\frac{\tilde{C}}{4\pi\sigma^{2}})(x_{0}^{2}+y_{0}^{2}) \Biggr]=0.
\end{align}
\end{subequations}
Equations \eqref{eq:EOM} demonstrate that the nonlinear rotation significantly influences the oscillation of the center of mass and the width of the condensate. It is also observed that the dynamical equations for the center of mass and width depend on $\dot{x_ 0}$ and $\dot{\sigma}$, respectively. This dependency indicates a coupling between the dynamical equations, which is not observed in the single component case~\cite{Pa:1997,Pa:1997b}. Consequently, the dynamics of the center of mass and the width will exhibit different behavior from those of single-component condensates due to the influence of nonlinear rotation.

To determine a mathematical expression for the dipole mode frequency, we consider the small amplitude oscillation of {$x_0, y_0$ and $\sigma$} around the equilibrium points. After linearizing Eqs. \ref{eq:EOM}(a)-(c) around the equilibrium point $x_0=0$, $y_0=0$, and $\sigma=\sigma_0$, where $\sigma_0$ is the width of the Gaussian ground state, we obtain
\begin{align}
\omega_{\rm d}=1-\left(\Omega+\frac{\tilde{C}}{4\pi\sigma_{0}^2}\right)
\end{align}
where the dipole mode frequency decreases with increasing $\tilde{C}$, the same phenomenon which was also found from the BdG spectrum. The dipole mode obtained using \me{the} variational approach is \me{shown as a} dashed line in Fig.~\ref{fig:comparison_com}(g). 
Additionally, the breathing mode is not illustrated in Fig.~\ref{fig:excit}. To identify this mode, we can perform a Fourier transform of the time evolution of the root mean square radius $r_{\rm rms}$ after \me{quenching the harmonic trapping strength}~\cite{Banger:2023}, similar to \me{the analysis of the} dipole mode. However, observations from the variational approach indicate that this mode is independent of the nonlinear rotation strength $\tilde{C}$. Therefore, the numerical calculation of the breathing mode is not performed here. {Proceeding, the} frequency of the breathing mode is obtained using variational approach as 
\begin{align} 
\omega_{\rm b} = \sqrt{\frac{\text{g}}{2\pi\sigma_0^4} - \frac{3}{\sigma_0^4} - 1 }
\label{eq:breathing} 
\end{align} 
Equation \eqref{eq:breathing} indicates that the breathing mode frequency remains unaffected by the density-dependent gauge potential. To determine the breathing frequency, we first compute the equilibrium condensate width $ \sigma_0 $ by solving the steady-state equation:
\begin{align} 
\sigma_0^4 - \frac{\text{g}}{2\pi\sigma_0^3} - 1 = 0, 
\end{align} 
which yields $ \sigma_0 = 2.83 $ for our parameter set. Substituting this value into Eq.~\eqref{eq:breathing}, we find the breathing mode frequency to be $ \omega_{\rm b} = 1.93 $. In Fig.~\ref{fig:comparison_com}(a), we present a comparison of {dipole mode} frequencies obtained using three distinct techniques. {Here} the frequencies derived from the perturbative method and the variational approach show excellent agreement with those from the {Bogoliubov-de Gennes} spectrum, demonstrating a strong agreement between these methods.

\section{Vortex-displacement modes}
\label{sec:vort-disModel}
 
In this section, we explore the \me{effect} of the density-dependent gauge potential \me{on the various} vortex displacement modes. One of these displacement modes, known as the Tkachenko mode, is represented by a green triangle ($\textcolor{green}{\triangledown}$) in the excitation spectrum {presented in Fig.~\ref{fig:excit}}  Within the rotating frame, these modes describe the motion of the vortex core around its equilibrium position. These can be categorized into four fundamental types, as discussed by Campbell~\cite{Campbell:1981} and Simula~\cite{Simula:2013}, and are described as follows: 
 \begin{enumerate}
 \item {\bf Tkachenko Mode (T):} The Tkachenko mode represents a fundamental vortex displacement mode characterized by a torsional oscillation of the vortex lattice, where each vortex executes an elliptical motion around its equilibrium position. It is one of the zero modes associated with the coupled \me{Eqs.}~\eqref{eq:bdg} and is related to the breaking of $\mathrm{SO}(2)$ symmetry. Unlike the usual orthonormalization conditions, such as $\int \left[u^{}_i(x,y) u_j(x,y) - v^{}{_i}(x,y) v_j(x,y)\right] d^2r = \delta{_{ij}}$ and $\int \left[u_i(x,y) v_j(x,y) - u_j(x,y) v_i(x,y)\right] d^2r = 0$, the Tkachenko mode satisfies \me{instead} the condition $\int u_1^{}(x,y) u_1(x,y) d^2r = \int v_1^{}(x,y) v_1(x,y) d^2r$ for \me{the appropriate} eigenstates. 
 \item \textbf{Common Mode (C):} This mode represents the center-of-mass excitation of the vortex array. The entire vortex lattice undergoes retrograde circular motion around the trap center as a rigid body in a frame {co-rotating} with the trapping potential.
 \item \textbf{Quadratic Modes (Q):} These modes involve a rigid-body oscillation of the central vortex while other surrounding vortices can move relative to one another.
 \item \textbf{Rational Modes (R):} In these modes, the central vortex remains stationary while the other vortices exhibit relative motion within the lattice.
 \end{enumerate}

\begin{figure}[!htp]
 \includegraphics[width=\linewidth]{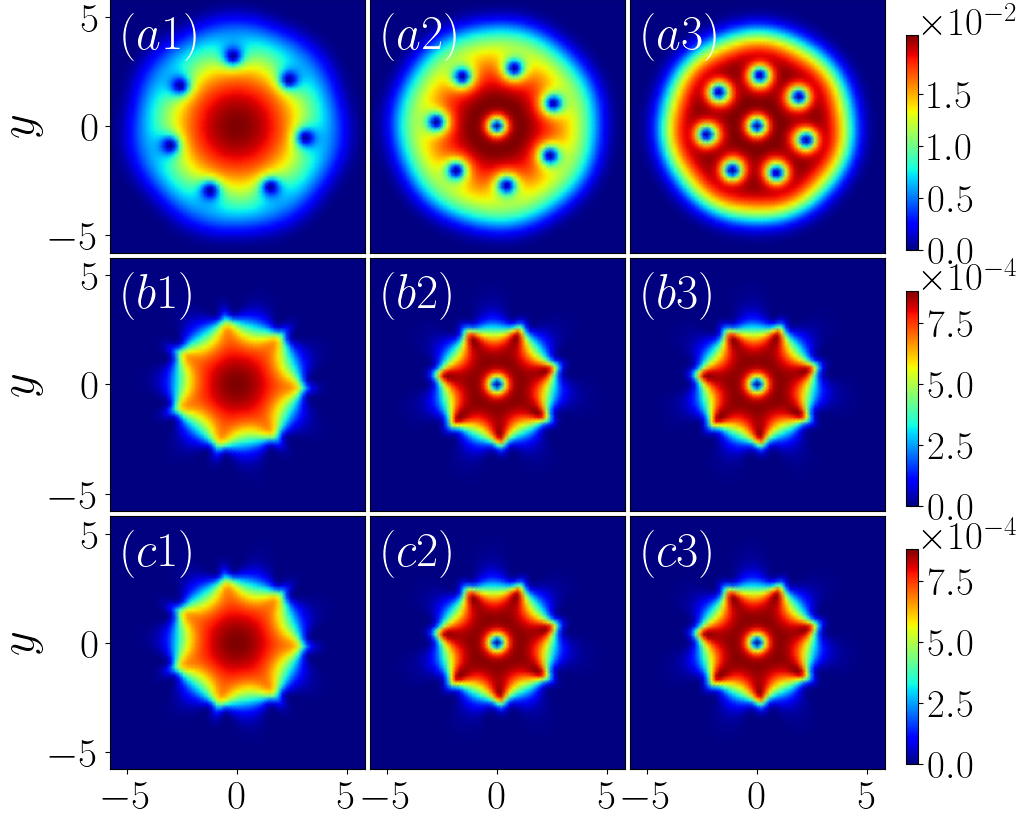}
 \caption{Pseudocolor representation of the condensate density for (a1) $\tilde{C} = -30$, (a2) $\tilde{C} = 0$, and (a3) $\tilde{C} = 20$. Panels (b1)–(b3), and (c1)-(c3) show the eigen mode functions $\lvert u(x, y) \rvert^2$, and $\lvert v(x, y) \rvert^2$ corresponding to the Tkachenko mode.}
 \label{fig:density_Tkachenko}
 \end{figure}

So far in this work we have investigated how the frequencies of the excitation modes are influenced by the nonlinear rotation induced by the gauge potential.
Next we move our focus \me{to} the higher-order modes by presenting the density distribution of the eigenfunctions associated with the Tkachenko mode. In Fig.~\ref{fig:density_Tkachenko} we illustrate $\lvert u(x, y)\rvert^{2}$ and $\lvert v(x, y) \rvert^{2}$ for specific values of $\tilde{C} = -30$, $0$, and $20$. Here, the eigenfrequencies \me{satisfy} the aforementioned condition $\lvert u(x, y)\rvert^{2} = \lvert v(x, y) \rvert^{2}$ [see middle and lower panel]. The density distribution reveals distinct peaks at the positions of the vortex cores. As a result, the perturbation described in Eq.~\eqref{eq:prtrb_den} shifts the vortex cores from their equilibrium positions, causing them to trace elliptical trajectories. 

In addition to the perturbation caused by the eigenfunctions obtained from the Bogoliubov-de Gennes analysis, the Tkachenko oscillation can also be excited experimentally through two different mechanisms. In one approach~\cite{Engles:2003}, a focused resonant laser beam removes atoms from the center of the condensate, creating a density dip. The condensate responds by flowing inward to fill the dip and outward to expand its radius. The Coriolis force then redirects the inward flow in the direction of the lattice rotation while the outward flow moves in the opposite direction. This shearing motion disturbs the vortices from their equilibrium positions and triggers oscillations in the lattice.\\
In the second method~\cite{Coddington:2003}, a resonant laser light is employed to generate a Gaussian dip in the radial trapping potential to produce an inward flow of the condensate that resembles the dynamics observed in the first method. Motivated by this second approach, we introduce a Gaussian perturbation of the form $V_{\rm p}(x,y)=-\exp\left(-(x^2+y^2)/(R_{TF}/2)^2\right)$, where $R_{\rm TF}=\sqrt{2\mu/m\omega_{\perp}^2}$ is the Thomas-Fermi radius to Eq.~(\ref{eq:dmless_eq}). 
{This leads to an effective potential defined as 
\begin{align}
V_{\rm eff}(x,y)=V(x,y) -\exp\left(-(x^2+y^2)/(R_{TF}/2)^2 \right),
\label{eqn:Veff}
\end{align}}
where $V(x,y)$ is the harmonic trapping potential. This allows us to investigate the nature of Tkachenko oscillations through the time evolution of {particular} physical observables \cite{Mizushima:2004}. We then evolve the previously obtained stationary state in real-time using the modified {Gross-Pitaevskii equation with Eq.~\eqref{eqn:Veff} and} analyze the resulting dynamics, which helps to validate the results obtained from the BdG analysis.

\begin{figure}[!htp]
 	\centering
 	\includegraphics[width=\linewidth]{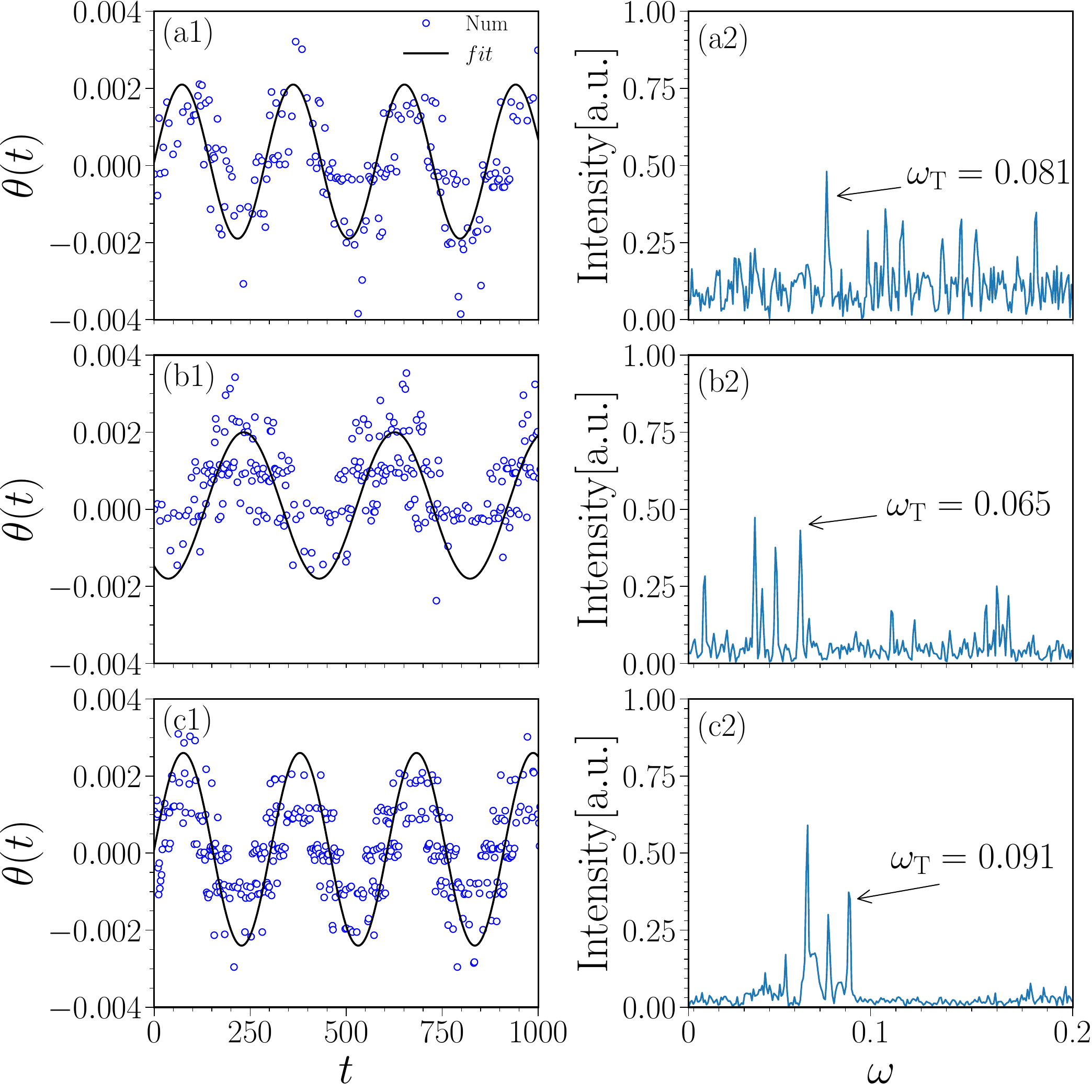}
 	\caption{Transverse oscillations of vortices for (a1) $\tilde{C} = -20$, (b1) $\tilde{C} = 0$, and (c1) $\tilde{C} = 20$. Panels (a2)-(c2) show the corresponding Fourier analysis of these oscillation patterns. The solid black lines represent sinusoidal fits of the form $A \sin(B\omega+D)$. The fitting parameters are: $A = 0.0025$, $-0.0019$, and $0.004$; $B = 0.0217$, $0.016$, and $0.0207$ for $\tilde{C} = -20$, $0$, and $20$, respectively. The phase shift $D$ is fixed at $0.1$.}
 	\label{fig:tkachenko_theta}
 \end{figure}
 \begin{figure}[!htp]
 	\centering
 	\includegraphics[width=\linewidth]{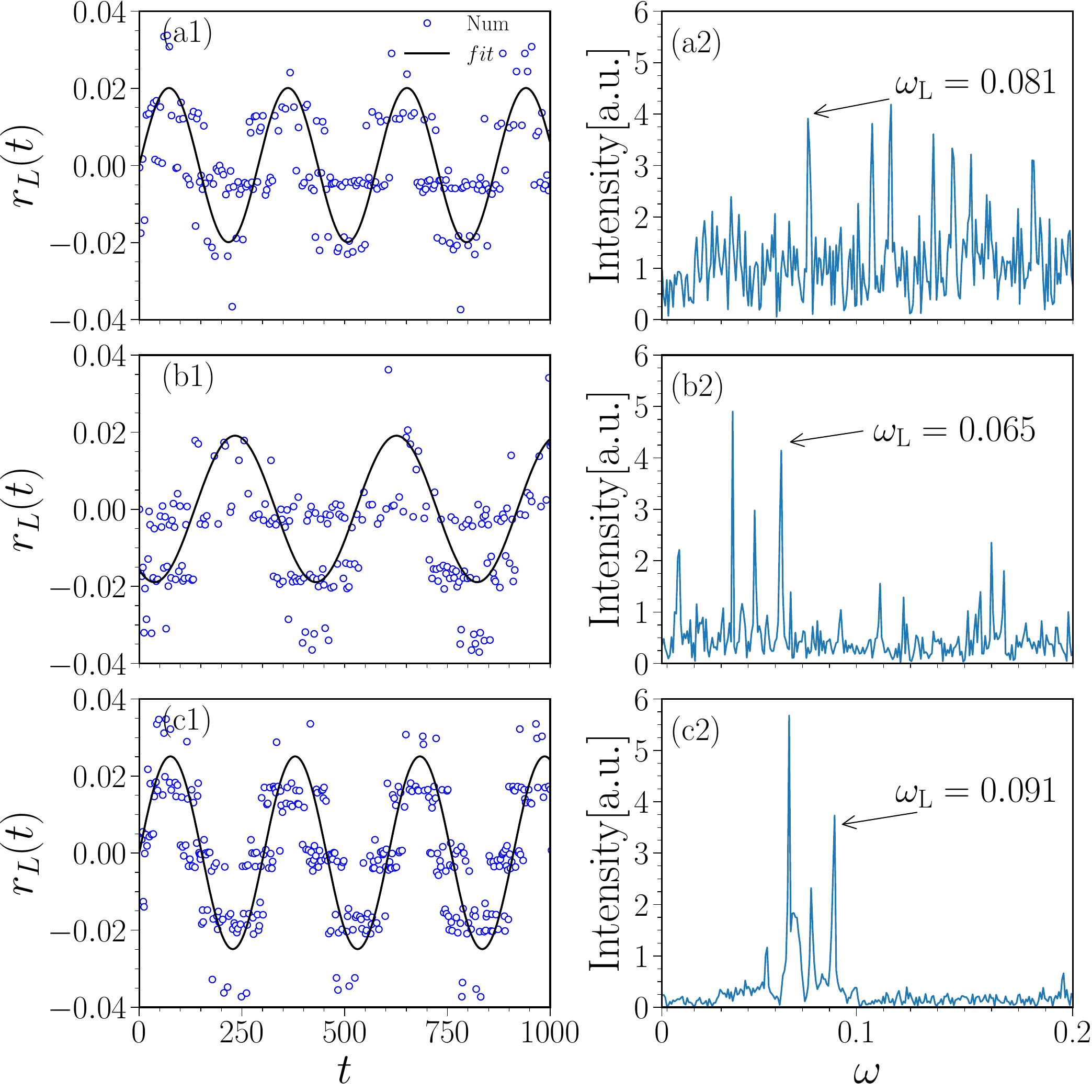}
 	\caption{Longitudinal oscillations of vortices for (a1) $\tilde{C} = -20$, (b1) $\tilde{C} = 0$, and (c1) $\tilde{C} = 20$. Panels (a2)-(c2) show the corresponding Fourier analysis of these oscillation patterns. The solid black lines represent sinusoidal fits of the form $A \sin(B\omega+D)$. The fitting parameters are: $A = 0.0025$, $-0.0019$, and $0.004$; $B = 0.0217$, $0.016$, and $0.0207$ for $\tilde{C} = -20$, $0$, and $20$, respectively. The phase shift $D$ is fixed at $0.1$.}
 	\label{fig:tkachenko_r}
 \end{figure}
Since the Tkachenko mode involves an elliptically polarized collective motion of vortices around their equilibrium positions, it can be decomposed into transverse and longitudinal polarization components. These correspond to the azimuthal and radial oscillations of the vortices, respectively.
The transverse vortex motion is characterized by the averaged angular displacement
\begin{align}
\theta(t) = \sum_{i=1}^{N_v} \left( \arctan\left(\frac{y_i(t)}{x_i(t)}\right) - \arctan\left(\frac{y_0(t)}{x_0(t)}\right) \right),
\end{align}
where $N_v$ represents the total number of vortices, excluding the central vortex.
The longitudinal vortex motion, representing the radial displacement of the vortices, is given by
$r_{L}(t) = \sum_{i=1}^{N_v} \lvert r_{i}(t) - r_{i}(0) \rvert$.

In Fig.~\ref{fig:tkachenko_theta}, we illustrate the temporal evolution of the transverse motion of vortices for three different values of $\tilde{C}$: $-20$, $0$, and $20$ (see Figs. (a1)-(c1)). We also compute the Fourier transform of the motion to determine the frequency of the Tkachenko mode. Our findings indicate that at $\tilde{C} = 0$, the Tkachenko mode frequency is $\omega_{T} = 0.065$. This frequency increases to $0.081$ for $\tilde{C} = -20$ and further to $0.089$ for $\tilde{C} = 20$.
In Fig.~\ref{fig:tkachenko_r}, we present the time evolution of the longitudinal motion of vortices, along with its Fourier spectrum. It is observed that the sharp peaks at $\omega_{L}=0.081, 0.065$, and $0.089$ coincide with $\omega_{T}$, indicating that the vortex core exhibits elliptical motion around its equilibrium position. The real-time evolution of Eq.~\eqref{eq:dmless_eq} requires a very small time step, $\Delta t \approx 10^{-7}$, due to the density-dependent rotation term. This significantly increases the computational cost. To reduce this cost, we introduce a small dissipation parameter, $\gamma = 2 \times 10^{-3}$, in Eq.~\eqref{eq:dmless_eq} by replacing $i$ with $(i - \gamma)$ on the left-hand side of the equation. 
In order to capture the temporal oscillation of the vortex more adequately we fit the transverse $\theta(t)$ and longitudinal motion ($r_{L}(t)$) of the vortices with a sinusoidal curve as denoted with a solid black line. The fitting suggests the sinusoidal nature of the vortex oscillation as the system is perturbed by the low-lying collective excitation modes. We have provided the corresponding time evolution \me{animations} as a supplementary \me{files} [see supplimentary movies]. This particular feature of the vortex dynamics may be attributed to the excitation of the Tkachenko mode as also reported for the rotating BECs~\cite{Mizushima:2004}. It should be noted that due to a structural change of the vortex lattice from the hexagonal configuration to a ring formation, the time series of \me{the} vortex evolution appears to \me{contain some noise} which also \me{becomes} manifested as an appearance of multiple peaks in the frequency spectrum. This also may be the reason that we detect fewer frequencies for $\tilde{C} \geq 0$~(see \ref{fig:tkachenko_theta}(b1)-(b2), and \ref{fig:tkachenko_theta}(c1)-(c2)) compared to those we find for $\tilde{C} <0 $~(see Fig.~\ref{fig:tkachenko_theta}(a1)-(a2)). One may decrease the noise in both the time series and frequency spectrum for $\tilde{C} \geq 0$ by considering a larger vortex lattice which can be attained \me{in the} higher \me{rigid body} frequency limit ($\Omega\sim 1$)~\cite{Mizushima:2004}. 


To gain a better understanding of the Tkachenko mode frequency, we derive its analytical expression using the continuum hydrodynamic approach. This method has been previously applied to describe the Tkachenko mode in both single~\cite{Baym:2003} and two-component~\cite{Kecceli:2006} BECs. Following the same framework, we obtain the following expression for the Tkachenko mode frequency:
\begin{align}
	\omega_{T}^2 = \frac{2C_2}{nm} \frac{s^2k^4}{(\Omega_{\text{vort}}^2 +
		(s^2+4(C_1+C_2)/nm)k^2)}.
	\label{eq:tkachenko}
\end{align}
Here, $C_1$ and $C_2$ represent the compressional and shear moduli of the vortex lattice, respectively, while $n$ denotes the density, $m$ is the mass of the superfluid, and $k$ is the wave vector. Additionally, $s$ corresponds to the sound velocity, and $\Omega_{\rm vort}$ represents the vorticity of the system. The detailed derivation of this expression is provided in {Appendix} \ref{appendix}. To determine the Tkachenko mode frequency from Eq.~\eqref{eq:tkachenko}, we evaluate the sound velocity at the trap center as $s_0 = \sqrt{\text{g} n_0}$, where $n_0 = \mu / \text{g}$ represents the central density. The effective wave vector is inversely proportional to the Thomas-Fermi radius, i.e., $k = \alpha / R_{\rm Th}$, and where $\alpha$ is a free parameter, which \me{we choose as} $\alpha=7$ in our work. The vorticity is computed using the expression 
\begin{align}\label{eqn:vort}
\mathbf{\Omega_{\text{vort}}} = \left[2\Omega + C \left( x \frac{\partial n(x,y)}{\partial x} + y \frac{\partial n(x,y)}{\partial y} \right)\right]\hat{e_{z}}
\end{align}
{here the} peak value {of $\mathbf{\Omega_{\text{vort}}}$} is used to determine the Tkachenko mode frequency.

\begin{figure}[!htp]
 	\centering
	\includegraphics[width=\linewidth]{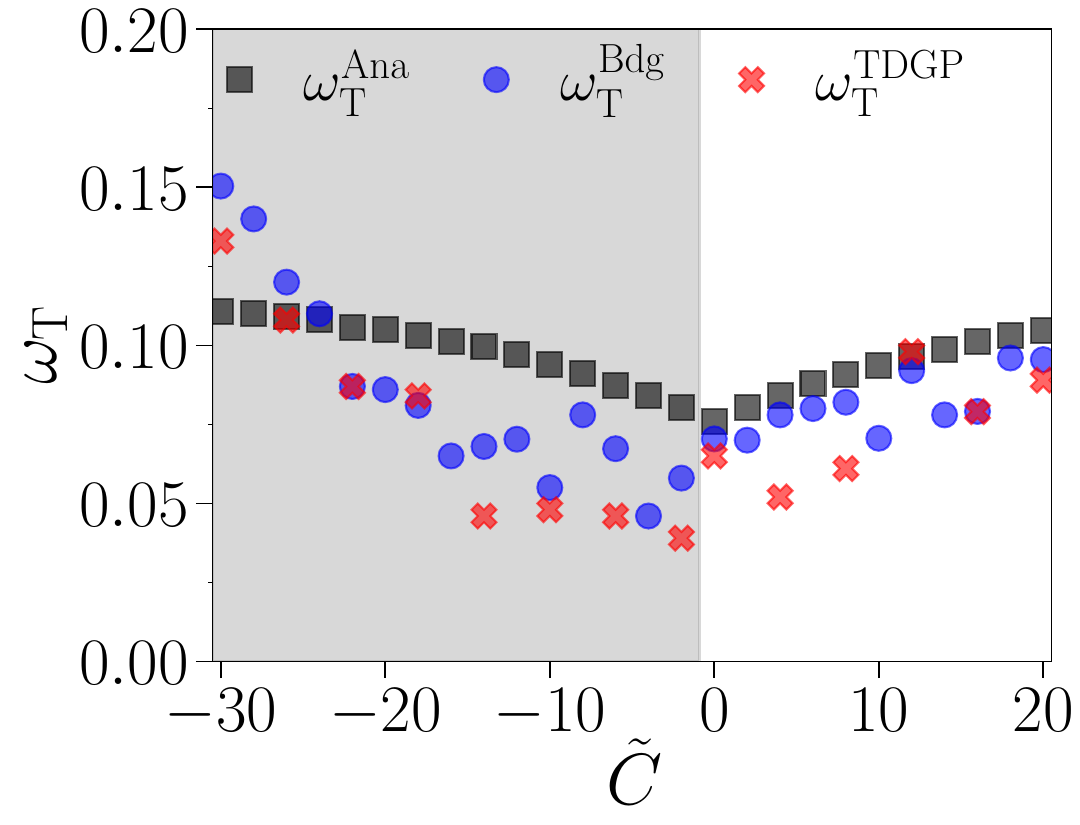}
	\caption{Variation of the Tkachenko mode frequency $ \omega_{T} $ as a function of nonlinear rotation $ \tilde{C} $. The filled black squares, blue circles, and crosses represent the Tkachenko frequency obtained from continuum hydrodynamic theory \me{(Eq.~\eqref{eq:tkachenko})}, BdG analysis, and the Gaussian perturbation applied to the time-dependent GPE, respectively.}
\label{fig:tkachenko}
\end{figure}

In Fig.~\ref{fig:tkachenko}, we present the variation of \me{the} Tkachenko mode frequency $\omega_{T}$ as a function \me{of the} nonlinear rotation $\tilde{C}$ \me{strength} obtained from three distinct techniques: BdG analysis, by perturbing \me{the} time-dependent GPE using Eq.~\eqref{eqn:Veff}, and the continuum hydrodynamic approach. We find that $\omega_{T}$ \me{has a} minimum around $\tilde{C} = 0$, and exhibits a gradual increase, following a quadratic dependency as $\tilde{C}$ deviates from zero in either the positive or negative direction. We observe a strong agreement between the values of $\omega_T$ obtained from \me{the} BdG analysis and the time-dependent Gross-Pitaevskii equation (TDGP). However, the frequency predicted by the hydrodynamic approach matches quite well for positive values of $\tilde{C}$, whereas for the negative range (shown by light grey region), the $\omega^{\rm Ana}_T$ (marked by black filled squares) deviates in compare to other approaches. This discrepancy arises from the underlying assumption of hydrodynamic continuum theory, which considers the vortex core size to be significantly smaller than the intervortex distance in the lattice. Here it is instructive to mention that for $\tilde{C} < 0$, the vortices in general get arranged on a ring at the periphery of the condensate and contain only a few number of vortices. As a result of this the assumption of continuum hydrodynamic theory appears not be valid for the region of $\tilde{C} < 0$ which may be linked to the reason for the apparent deviations observed between the Bogoliubov modes and the predictions from the continuum theory. However \me{for} $\tilde{C} > 0$ \me{the} Magnus force generated from the density-dependent gauge potential \me{causes} the vortices in general to \me{become} located near the center of the trap that results in \me{an} overall reduction in the inter-vortex spacing. 

So far, we have analyzed vortex dispacement modes by studying the post perturbation dynamics of \me{the} condensate density for different nonlinear rotation \me{strengths}. In Table~\ref{table:mode} we tabulate the mode frequencies of different vortex displacement modes across various nonlinear rotation strengths. For negative values, we examine $\tilde{C} = -30, -20, -12, -4$, and for positive values, we consider $\tilde{C} = 0, 4, 8, 12, 20$. This comprehensive overview indicates that the frequency of the vortex displacement mode is sensitive to $\tilde{C}$. It is important to note that neither the {dipole} mode (D) nor the surface mode (S) is responsible for the vortex displacement. The surface mode {consists of} surface excitations and nucleates additional vortices at the edge of the condensate. Furthermore, it has been observed that the energy of the surface mode remains relatively low, whereas it increases as $\tilde{C}$ becomes positive. For example, as shown in Table~\ref{table:mode}, the surface mode energy is $\omega=0.561$ at $\tilde{C}=-30$, and rises to $\omega=0.769$ when $\tilde{C}=20$. This phenomenon \me{occurs} because for $\tilde{C}<0$, the condensate density at the edge decreases, making it energetically favorable to form the surface ripples. However, for $\tilde{C}>0$, the density profile shows a large plateau region, that requires more energy to create such ripples. We have observed that the rational mode (R) is present for $\tilde{C} = -4$ and $\tilde{C} = -12$; however, it is absent for both higher negative values and any positive values of $\tilde{C}$. For large negative values, vortices are clustered along the periphery of the condensate. In contrast, with positive $\tilde{C}$, the vortices become concentrated near the trap \me{center}. In either scenario, the hexagonal symmetry is broken, highlighting the influence of $\tilde{C}$ on the vortex displacement modes.
\begin{table}[!t]
	\begin{center}
		\begin{tabular}{|c|c|c|c|cccc|c|c|c|}
			\hline\hline
			$\tilde{C}$ & mode & $\omega$ & label &&&&$\tilde{C}$ & mode & $\omega$ & label\\
			\hline
			& 1 & 0.023 & C &&&& & 1 & 0.088 & C \\
			
		 & 2	& 0.150	& T &&&& & 2 & 0.078 & T \\
			 
			 & 3 & 0.263 & Q &&&&	 & 3 & 0.105 & Q \\
			 
			-30 & 4 & 0.561 & S &&&& 4 & 4 & 0.181 & Q \\
			 
			 & 5 & 0.696 & D &&&& & 5 & 0.285 & Q \\
			 
			 & 6 & 0.785 & S &&&& & 6 & 0.417 & D \\
			 
			 & 7 & 0.827 & S &&&& & 7 & 0.661 & S \\
			 \hline

			 & 1 & 0.019 & C &&&& & 1 & 0.059 & C \\ 
		
	 	 & 2	& 0.087	& T &&&& & 2 & 0.084 & T \\ 
		
		 & 3 & 0.152 & Q &&&& & 3 & 0.107 & Q \\
		
		 -20 & 4 & 0.294 & Q &&&& 8 & 4 & 0.176 & Q \\
		
		 & 5 & 0.500 & D &&&& & 5 & 0.261 & Q \\
		 
		 & 6 & 0.544 & S &&&& & 6 & 0.361 & D \\
		 
	 & 7 & 0.588 & S &&&& & 7 & 0.629 & S \\ 
		
		\hline
		
		 & 1 & 0.018 & C &&&& & 1 & 0.094 & C \\
		
		& 2	& 0.0703	& T &&&& & 2 & 0.092 & T\\
		
		& 3 & 0.113 & R &&&& & 3 & 0.154 & Q\\ 
		
	-12	& 4 & 0.153 & Q &&&& 12 & 4 & 0.216 & Q \\
		
		& 5 & 0.298 & Q &&&& & 5 & 0.271 & Q \\
		& 6 & 0.439 & D &&&& & 6 & 0.351 & D \\
		
		& 7 & 0.667 & S &&&& & 7 & 0.763 & S \\
		
		\hline
		
	 & 1 & 0.040 & C &&&& & 1 & 0.079 & T \\
		
		& 2	& 0.046	& T &&&& & 2 & 0.086 & C \\
		
		& 3 & 0.158 & R &&&& & 3 & 0.228 & Q \\
		 
	-4	& 4 & 0.226 & Q &&&& 16 & 4 & 0.280 & Q \\
		
		& 5 & 0.339 & Q &&&& & 5 & 0.316 & D \\
		& 6 & 0.425 & D &&&& & 6 & 0.739 & S \\
		
		& 7 & 0.665 & S &&&& & 7 & 0.818 & S \\
		
 \hline
		
	 & 1 & 0.067 & C &&&& & 1 & 0.085 & C \\
		
		& 2	& 0.070	& T &&&& & 2 & 0.096 & T\\

		& 3 & 0.120 & Q &&&& & 3 & 0.171 & Q\\
		
	0	& 4 & 0.170 & Q &&&& 20 & 4 & 0.23 & D \\
		
		& 5 & 0.281 & Q &&&& & 5 & 0.283 & Q\\
		& 6 & 0.436 & D &&&& & 6 & 0.450 & Q \\
		& 7 & 0.676 & S &&&& & 7 & 0.769 & S \\
			
			\hline
		\end{tabular}
	\end{center}
	\caption{Frequencies of the low-lying excitations for $\Omega = 0.6, g = 400$ at different values of $\tilde{C}$. Here, the notations represents the circular (C), dipole (D), Quadrupole (Q), Tkachenko (T), surface (S), and rational (R) modes. Movies illustrating the dynamical behavior of all listed modes are available in the Supplemental Material.}
	\label{table:mode}
\end{table}


\begin{figure}[!htp]
 \centering
 \includegraphics[width=\linewidth]{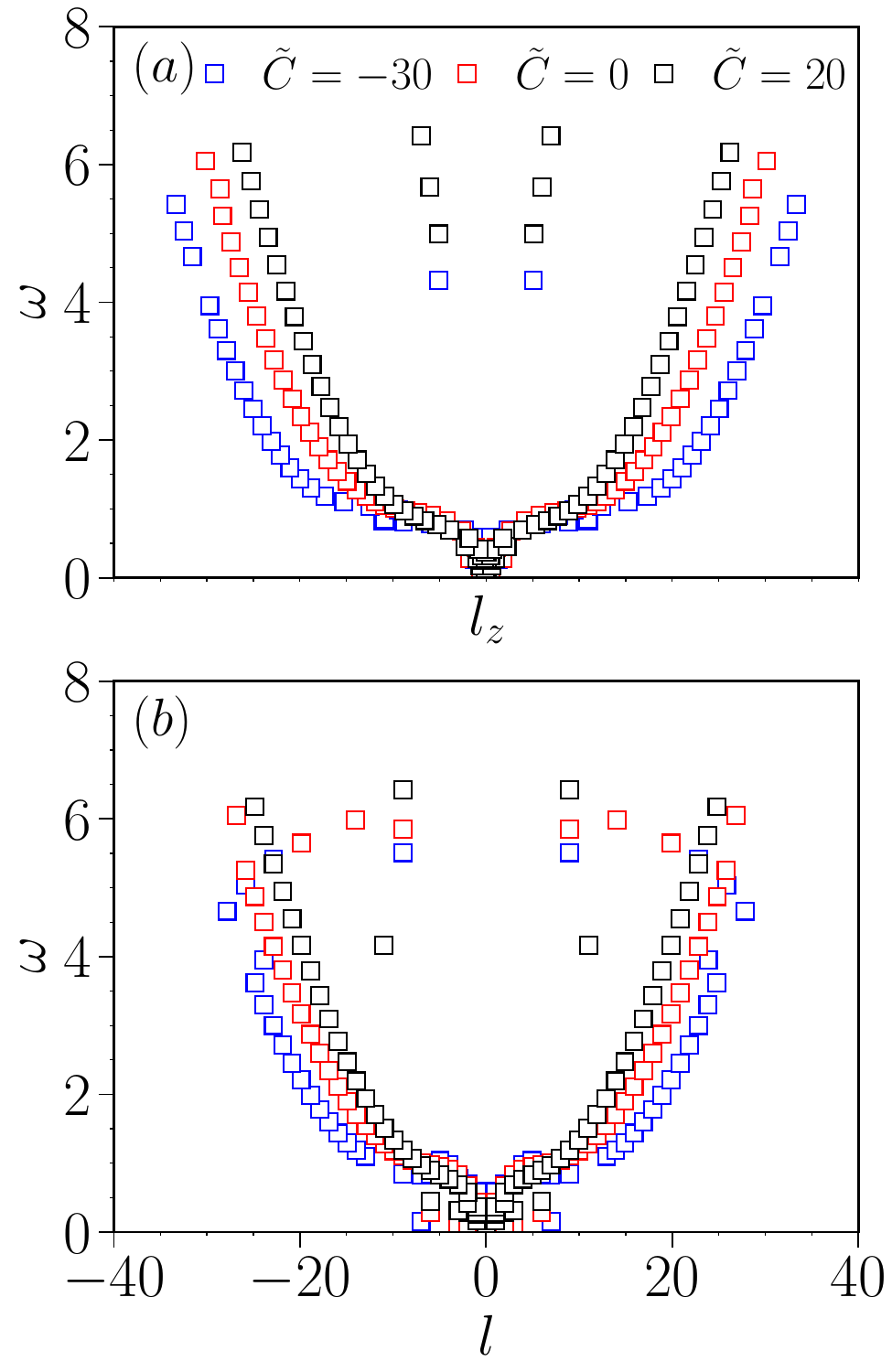}
 \caption{Excitation spectrum as a function of (a) angular momentum, and (b) angular quantum number for $\tilde{C}=-30,0$, and $20$.}
 \label{fig:ang_vs_omg}
 \end{figure}
\begin{figure}[!htp]
 \centering
 \includegraphics[width=0.9\linewidth]{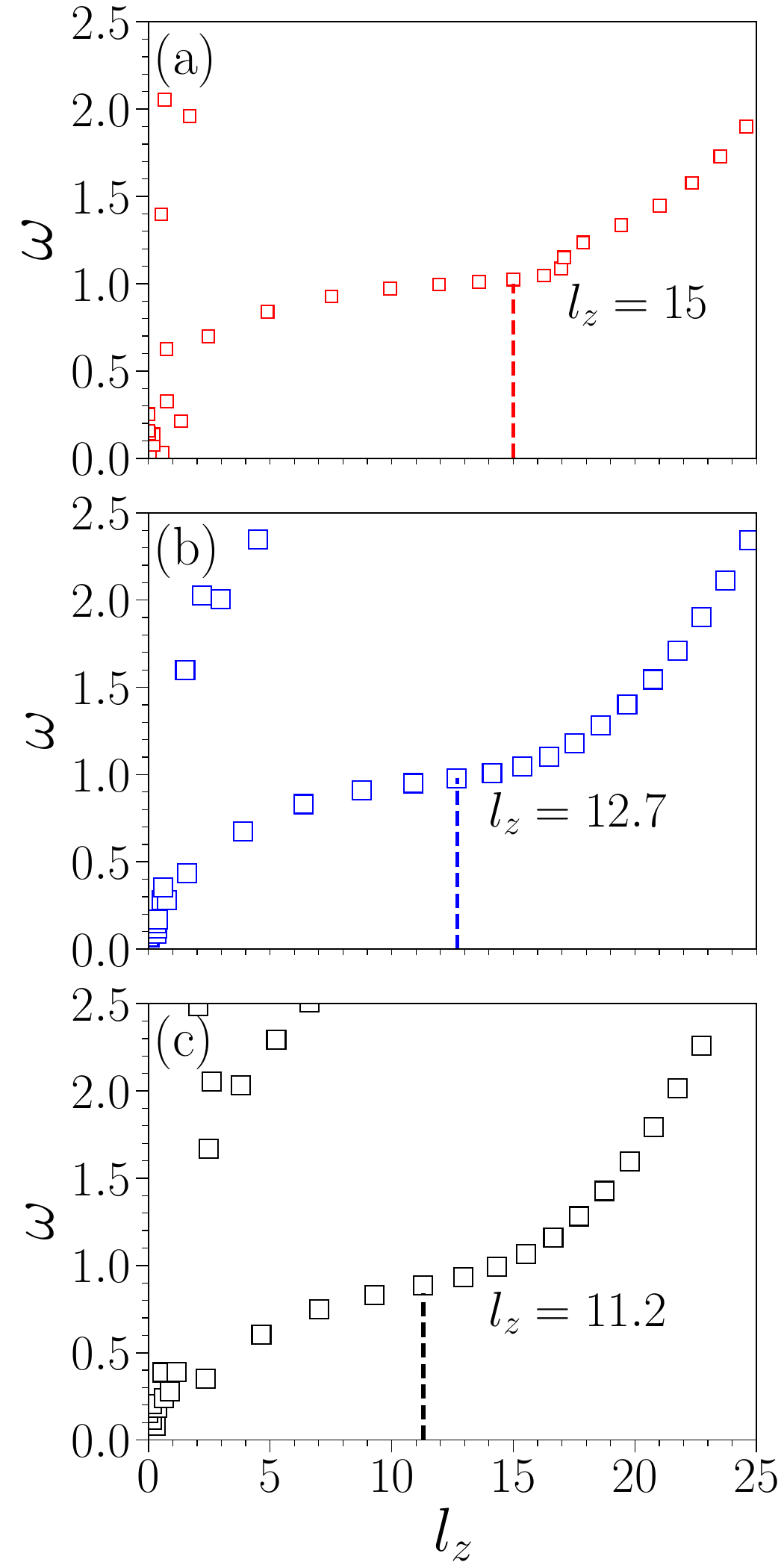}
 \caption{The frequencies of the low-lying excitations are shown as a function of their orbital angular momentum relative to the condensate ground state for (a) $\tilde{C} = -30$, (b) $\tilde{C} = 0$, and (c) $\tilde{C} = 20$. For $\tilde{C} = -30$, the roton-like minima shift toward higher angular momentum, and towards lower angular momentum for $\tilde{C} = 20$.
}
 \label{fig:ang_vs_omg_minima}
 \end{figure}

Figure \ref{fig:ang_vs_omg}(a) and (b) show the excitation spectrum $\omega$ as a function of angular momentum $l_z$ and the angular quantum number $l$, respectively, for three different values of $\tilde{C}$: $\tilde{C} = -30$, $0$, and $20$. The angular momentum of the excitations relative to the rotating condensate is defined as \cite{Woo:2004} 
\begin{align}
l_z=\frac{(l_u - l_\psi)N_u + (l_v + l_\psi)N_v}{N_u + N_v},
\end{align}
where 
\begin{align}
l_\alpha=-i\hbar\frac{\int dxdy\alpha^*\big(x \frac{\partial\alpha}{\partial y} - y\frac{\partial\alpha}{\partial x}\big)}{\int dxdy\alpha^*\alpha},
\end{align}
with $\alpha$ being one of the wave functions $u(x,y)$, $v(x,y)$, or $\psi(x,y)$, and $N_u = \int dxdy |u(x,y)|^2, \quad N_v = \int dxdy |v(x,y)|^2$. To determine the angular quantum number $l$, we define $n^{\prime}=\psi^{*}(x,y)u(x,y)-\psi(x,y)v(x,y)$. The angular quantum number $l$ is then determined by evaluating the $\nabla S$ along a closed path that encircles the condensate at the Thomas-Fermi radius: $\frac{1}{2\pi} \oint_C \nabla S \, \text{d}r$, where $S$ is the phase of $n^{\prime}$. The excitation spectrum shows a high degree of symmetry around $l = 0$. Notably, the energy splitting between the modes with angular momentum $l$ and $-l$ is influenced by the nonlinear rotation parameter $\tilde{C}$. Both Figures \ref{fig:ang_vs_omg}(a) and (b) illustrate that the energy splitting, for both angular momentum and angular quantum number, is larger when $\tilde{C} = -30$ and diminishes substantially when $\tilde{C} = 20$. It has also been observed that the energy of the surface excitation of the condensate increases for $\tilde{C}=20$, while it decreases for $\tilde{C}=-30$. In the excitation spectrum, we observe flattening of the energy below a threshold value of the angular momentum~(see Fig.~\ref{fig:ang_vs_omg_minima}) which may be attributed to the roton like energy dip observed for the low number of vortices as reported in~\cite{Simula:2013}. These roton-like minima may lead to the formation of additional vortices in the system by triggering instabilities at the surface of the condensate. We note that these hanging modes shift to higher angular momenta for negative values of $\tilde{C}$ and to lower angular momenta for positive values of $\tilde{C}$. This type of shifting is a direct consequence of the expansion and contraction of the Thomas-Fermi radius, which increases for negative $\tilde{C}$ and decreases for positive $\tilde{C}$.
\section{Conclusion} \label{sec:conclusion} 
In this work, we have investigated the effects of nonlinear rotation in a quasi-two-dimensional Bose-Einstein condensate, which arises from a density-dependent gauge potential on the collective excitation spectrum using Bogoliubov–de Gennes theory. We found that the strength of the nonlinear rotation parameter \me{influences} the excitation spectrum in several ways. By analyzing the time evolution of the condensate density profile, perturbed by the eigenstates of the collective mode, we identified different low-lying collective modes. 

Using a variational analysis, we have found that the center of mass oscillation and the width of the condensate are coupled with each other and also depend upon the strength of the nonlinear rotation. We also derived an analytical expression for the dipole and breathing \me{modes} using this approach. Additionally, we compared the frequency of the {dipole} mode by examining the time evolution of center of mass and incorporating an appropriate perturbation into the time-dependent Gross-Pitaevskii equation. Our findings indicate that the nonlinear rotation leads to a violation of the Kohn's theorem, where the frequency of the {dipole} mode depends on the interaction strength. 
We have shown that the frequency of the Tkachenko mode strongly depends on the strength of the nonlinear rotation, and the frequency attains minimum at $\tilde{C}=0$, which symmetrically increases for both negative and positive values of $\tilde{C}$. Additionally, we also derived an analytical expression for the Tkachenko mode frequency using a continuum hydrodynamic approach which \me{was found to agree with our numerical results}. Furthermore, the frequency is comprehended through the time evolution of the transverse and longitudinal motion of the vortices when a Gaussian perturbation is applied to the Gross-Pitaevskii equation (GPE). In addition to the Tkachenko mode, we also identified other fundamental modes depending upon nonlinear rotation. Also, the excitation spectrum \me{was} presented as a function of \me{the} angular momentum, \me{showing} that the energy splitting between $l$ and $-l$ depends on the strength of the nonlinear rotation. Subsequently, the presence of roton-like \me{minima} in the excitation spectrum explained the presence of surface mode \me{excitations} which shift to higher angular momentum for negative values of nonlinear rotation, \me{and} to lower angular momentum for positive values of nonlinear rotation.

The unusual nature of the superfluidity presented in this study presents opportunities for future studies related to synthetic nonlinear gauge potentials. It would be interesting to investigate how altering the confining geometry changes the vortex morphology \cite{Dubessy:2012,Woo:2012,Adhikari:2019}; moreover how the density-dependent gauge potential modifies the three-dimensional vortex line and ring solutions and their associated dynamics is an another avenue to potentially explore. Finally, the role of superfluid turbulence in this system provides a route to understanding the interplay of how energy is distributed across length scales with a spatially varying vorticity \cite{Tsubota:2017,Tsatsos:2016,White:2014}.

\acknowledgments
We thank Muneto Nitta and Patrik \"Ohberg for discussions.
R.B. acknowledges the financial support from the Department of Science and Technology's Innovation in Science Pursuit for Inspired Research (DST-INSPIRE) program, India. M.E.'s research was supported by the Australian Research Council Centre of Excellence in Future Low-Energy Electronics Technologies (Project No. CE170100039) and funded by the Australian government, and by the Japan Society of Promotion of Science Grant-in-Aid for Scientific Research (KAKENHI Grant No. JP20K14376). The work of P.M. is supported by the Ministry of Education-Rashtriya Uchchatar Shiksha Abhiyan (MoE RUSA 2.0): Bharathidasan University—Physical Sciences.

\appendix
\counterwithin{figure}{section}
\section{Derivation of the Tkachenko mode frequency using continuum hydrodynamic approach}
\label{appendix}
Here we present the detailed derivation of the Tkachenko mode frequency using the hydrodynamic approach. The superfluid acceleration equation for a condensate with velocity $\mathbf{v}$, vortex displacement field $\bm{\epsilon}$, and vorticity $\mathbf{\Omega_{vort}}$, can be expressed as~\cite{Baym:2003} ,
\begin{align}
	m\left(\frac{\partial \mathbf{v}}{\partial t} + \mathbf{\Omega_{\text{vort}}} \,\times
	\dot{\bm{\epsilon}}\right) = - \nabla \mu,
	\label{supaccel}
\end{align}
where $\mu$ is the chemical potential of the condensate. The elastic energy density of a triangular lattice in two dimensions is expressed in the following form:~\cite{Baym:2003},
\begin{align}
	{\cal E} = 2C_1 (\nabla\cdot\epsilon)^2
	+C_2\left[\left(\frac{\partial \epsilon_x}{\partial x}
	-\frac{\partial\epsilon_y}{\partial y}\right)^2
	\right. \notag \\ \left.
	+ \left(\frac{\partial \epsilon_x}{\partial y} +\frac{\partial
		\epsilon_y}{\partial x}\right)^2\right],
	\label{elastic}
\end{align}
where $C_1$ is the compressional modulus, and $C_2$ the shear modulus of the vortex lattice, and in terms of vorticity for an incompressible fluid, $C_2= -C_1$, {and} 
\begin{align}
	C_2 =\frac{n\Omega_{\rm vort}}{16}.
	\label{C2drop}
\end{align}
 We use the peak value of vorticity~$\mathbf{\Omega_{\rm vort}}$ {defined by Eq.~\eqref{eqn:vort} of the text} to determine $C_2$, where $n$ is the density of the condensate. The conservation of particle number and momentum results in the following continuity and momentum conservation equations.
\begin{align}
	\frac{\partial n}{\partial t} + \nabla \cdot \mathbf{j} = 0,
	\label{contin}
\end{align}
 where $\mathbf{j}=n\mathbf{v}$ is the particle current. 
\begin{align}
	m\left(\frac{\partial \mathbf{j}}{\partial t} + \mathbf{\Omega_{\text{vort}}} \times \mathbf{j}\right) +\nabla P = -\bm{\sigma},
	\label{momcons}
\end{align}
where $P$ represents pressure, which is related to the chemical potential as $\nabla P = n\nabla \mu$. The stress vector $\bm{\sigma}$ can be derived from the total elastic energy~$E_{\rm el} =\int d^2r {\cal E}(r)$, given by 
\begin{align}
	\bm{\sigma} = \frac{\delta E_{\rm el}}{\delta \bm{\epsilon}}
	= -4C_1\nabla (\nabla\cdot \bm{\epsilon}) -2C_2\nabla^2 \bm{\epsilon}.
\end{align}
From Eqs. (\ref{momcons}) and (\ref{supaccel}), along with $\nabla P = n\nabla\mu$, we derive the relationship 
\begin{align}
\boldsymbol{\Omega}_{\text{vort}} \times (\dot{\boldsymbol{\epsilon}} - \mathbf{v}) = \frac{\boldsymbol{\nabla} \cdot \boldsymbol{\sigma}}{m n}.
\end{align}
The curl and divergence of this equation yield
\begin{align}
\nabla \cdot (\dot{\boldsymbol{\epsilon}} - \mathbf{v}) 
= \frac{\nabla \times \boldsymbol{\sigma}}{\boldsymbol{\Omega}_{\text{vort}} \, n m} 
= \frac{C_2}{\boldsymbol{\Omega}_{\text{vort}} \, n m} \nabla^2 (\nabla \times \boldsymbol{\epsilon}),
\label{diveps}
\end{align}
and 
\begin{align}
\nabla \times \dot{\bm{\epsilon}} + 
\mathbf{\Omega}_{\text{vort}} \, \nabla \cdot \bm{\epsilon} 
= -\frac{\nabla \cdot \bm{\sigma}}{\mathbf{\Omega}_{\text{vort}} \, n m} 
= \frac{C_2 + 2C_1}{\mathbf{\Omega}_{\text{vort}} \, n m} \nabla^2 (\nabla \cdot \bm{\epsilon}).
\label{xeps}
\end{align}
The divergence of the superfluid acceleration, {Eq.~\eqref{supaccel} is} 
\begin{align}
\left( -\frac{\partial^2}{\partial t^2} + s^2 \nabla^2 \right) n =
n\, \mathbf{\Omega}_{\text{vort}} \cdot (\nabla \times \dot{\bm{\epsilon}}).
\label{densosc}
\end{align}
The speed of sound $s$ is given by $m s^2 = \me{\partial P/\partial n}$~\cite{Baym:2003}. Then by using Eq.~(\ref{diveps}) and eliminating $\nabla \times \bm{\epsilon}$ while disregarding the terms of order $\nabla^4$, we derive the following equation:
\begin{align}
	\left(-\frac{\partial^2}{\partial t^2} + \frac{2C_2}{mn}\nabla^2\right)
	\nabla\cdot\bm{\epsilon} = \frac{1}{n}\frac{\partial^2}{\partial
		t^2} n.
	\label{epsosc}
\end{align}
After linearising the coupled equations (\ref{densosc}) and (\ref{epsosc}), we obtain the mode frequencies by solving the corresponding secular equation for a given wavevector $k$
\begin{align}
	\lambda(k,\omega) = &\omega^4 - \omega^2\left[\Omega_{\text{vort}}^2 + \left(s^2
	+\frac{4}{nm}(C_1+C_2)\right)k^2\right] \notag \\
	 &+ \frac{2s^2C_2}{nm}k^4 =0.
	\label{eq:eigen}
\end{align}
Equation~\eqref{eq:eigen} can be simplified to 
\begin{align}
\omega^4 - A\omega^2 + \frac{2s^2C_2}{nm}k^4 =0.
	\label{eq:simp}
\end{align}
where $A=\left[\Omega_{\text{vort}}^2 + \left(s^2+\frac{4}{nm}(C_1+C_2)\right)k^2\right]$.
By solving Eq.~\eqref{eq:simp} {for $\omega^2$} we obtain 
\begin{align}
 \omega^2=\frac{A}{2}\left[A \pm \sqrt{A^2-\frac{8s^2C_2k^4}{nm}}\right].
	\label{eq:eigen_soln}
\end{align}
For $2s^2C_2k^4/nm \ll A$ the mode frequencies are given by
\begin{align}
	\omega_I^2 = \Omega_{\text{vort}}^2 + \left(s^2+\frac{4(C_1+C_2)}{nm}\right)k^2,
	\label{inertial}
\end{align}
and
\begin{align}
	\omega_{T}^2 = \frac{2C_2}{nm} \frac{s^2k^4}{(\Omega_{\text{vort}}^2 +
		(s^2+4(C_1+C_2)/nm)k^2)}.
	\label{tk}
\end{align}
The frequency $\omega_{I}$ corresponds to the inertial mode of the condensate, which arises from the Coriolis force in the rotating frame, while $\omega_{T}$ denotes the Tkachenko mode. 



\newpage
\bibliography{reference}
\end{document}